\documentclass[useAMS,usenatbib]{mn2e}

\usepackage{times}
\usepackage{graphicx}
\usepackage{amsmath}
\usepackage{mathabx}
\usepackage{subfigure}
\usepackage{natbib}
\usepackage{txfonts}
\usepackage{journals}

\title[3D extinction map of GAC]
{A three dimensional extinction map of the Galactic Anticentre from
multi-band photometry}
\author[B.Q. Chen et al.]
{B.-Q. Chen,$^1$\thanks{E-mail:
bchen@pku.edu.cn (BQC); x.liu@pku.edu.cn (XWL).}\thanks{LAMOST Fellow.}
 X.-W. Liu,$^{1,2}$\footnotemark[1]
 H.-B. Yuan,$^2$\footnotemark[2]
 H.-H. Zhang,$^1$
 M. Schultheis,$^3$
\newauthor 
 B.-W. Jiang,$^4$
 Y. Huang,$^1$
 M.-S. Xiang,$^1$
 H.-B. Zhao,$^5$
 J.-S. Yao,$^5$
 and H. Lu$^5$
\\
$^{1}$Department of Astronomy, Peking University, Beijing 100871, P.\,R.\,China\\
$^{2}$Kavli Institute for Astronomy and Astrophysics,
Peking University, Beijing 100871, P.\,R.\,China\\
$^{3}$Universite de Nice Sophia-Antipolis, CNRS, Observatoire de Cote d'Azur, Laboratoire
Cassiopee, 06304 Nice Cedex 4, France\\
$^{4}$Department of Astronomy, Beijing Normal University, Beijing 100875, P.\,R.\,China\\
$^{5}$Purple Mountain Observatory, Chinese Academy of Sciences, Nanjing 21008, Nanjing,
P.\,R.\,China\\
 }

\begin{document}

\date{Accepted ???. Received ???; in original form ???}

\pagerange{\pageref{firstpage}--\pageref{lastpage}} \pubyear{2014}

\maketitle

\label{firstpage}

\begin{abstract}

We present a three dimensional
extinction map in $r$ band. The map has a spatial angular
resolution, depending on latitude,  between 3 -- 9\,arcmin
and covers the entire XSTPS-GAC survey area of
over 6,000\,$\rm deg^2$ for Galactic longitude
$\rm 140 \leq$ $l$ $ \leq 220\deg$ and latitude $\rm -40\leq$ $b$ $ \leq 40\deg$.
By cross-matching the photometric catalog of 
the Xuyi Schmidt Telescope Photometric Survey of the Galactic Anticentre (XSTPS-GAC)
with those of 2MASS and WISE,
we have built a multi-band photometric stellar sample
of about 30 million stars and
applied spectral energy distribution (SED) fitting to the sample.
By combining photometric data from the optical to the near-infrared,
we are able to break the degeneracy between the intrinsic stellar colours and
the amounts of extinction by dust grains
 for stars with high photometric accuracy,
and trace the extinction as a function of distance for low
Galactic latitude and thus highly extincted
regions. This has allowed us to derive the
best-fit extinction and distance information of more than
13 million stars, which are used to construct the
three dimensional extinction map.
We have also applied a 
Rayleigh-Jeans colour excess (RJCE) method to the data using the 2MASS and WISE
colour $(H-W2)$. The resulting RJCE extinction map is consistent with the integrated 
two dimensional map deduced using the  best-fit SED algorithm. However for individual stars, the
amounts of extinction yielded by the RJCE method suffer from larger
errors than those given by the  best-fit SED algorithm. 
\end{abstract}


\section{Introduction}

The Milky Way, our own Galaxy, is an archetypical disk galaxy and the only 
grand-design (barred) 
spiral for which the individual constituent stars can be 
spatially resolved and studied multi-dimensionally (in 
not only three dimensional position  and velocity,  and 
in chemical composition, but also in age, size and surface gravity). 
A major objective of the current astrophysical study is to 
understand how galaxies assemble and acquire their characteristic structure and properties.
Modern large scale surveys, represented by the extremely successful
Sloan Digital Sky Survey 
(SDSS; \citealt{York2000}), have revolutionized our understanding of  galaxy formation 
and evolution, including the still on-going assemblage process of the Milky Way.
The on-going LAMOST Spectroscopic Survey of the Galactic Anticentre 
(LSS-GAC; \citealt{Liu2013}), a major 
component of the Galactic surveys with LAMOST
(also known as the Guoshoujing Telescope),
 offers a unique opportunity to reveal the 
true multi-dimensional structure
of the Milky Way and address fundamental questions with regard to 
 the formation and evolution of 
the Galactic disk, and of the Galaxy as a whole. 

Based on the SDSS photometric data, a series of work have attempted to study
in detail the distribution of tens of millions of
Galactic stars in multi-dimensional space 
(e.g. \citealt{Juric2008,Ivezic2008,
Bond2010,Berry2011}). 
Most of those studies concentrate on regions of 
 high Galactic latitudes
 ($|b|$ $>$ 30$^{\circ}$). Although the Sloan Extension
for Galactic Understanding and Exploration (SEGUE; 
\citealt{Yanny2009}) has reached the
low latitude disk, it covers only a few stripes crossing the Galactic plane.  
In order to provide an input catalog for the LSS-GAC, 
a multi-band CCD photometric survey of the Galactic 
Anticentre with the Xuyi 1.04/1.20m Schmidt Telescope  
(XSTPS-GAC; \citealt{Zhang2013,Zhang2014,Liu2013}) 
has been carried out. The XSTPS-GAC photometric catalog contains 
more than 100 million stars in the direction of Galactic
anticentre (GAC), thus providing an  excellent 
data set  to study the Galactic disk, in particular the  
structures and substructures of the outer disk (truncation, warps and flares), and the
Monoceros Ring and other stellar (sub)structures in the GAC direction. The study 
is however challenged by the presence of a significant 
amount of prevailing dust extinction that varies
on small scales in the GAC direction.

Interstellar extinction is a serious obstacle for the interpretation of stellar populations 
in the Galaxy, especially for the low latitude regions, specially the Galactic disk. 
It shows an inhomogeneous clumpy  distribution  and increases towards the Galactic 
plane \citep{Burstein1982, Chen1998, Schultheis1999, Gonzalez2012, Nidever2012}. 
The currently available extinction maps of the GAC regions have various spatial resolutions 
and coverages. The most commonly used is the whole sky map of 
Schlegel et al.\,(1998, SFD hereafter). The SFD map is based on the
distribution of dust temperatures derived from the Infrared Astronomical Satellite 
(IRAS; \citealt{Neugebauer1984}) and the Diffuse Infrared Background Experiment (DIRBE;
\citealt{Boggess1992}) 
experiments. The colour temperatures are then calibrated 
 to values of colour excess $E(B-V)$ of dust reddening using
colours of background galaxies. Unfortunately, the map suffers from large 
uncertainties in regions towards the Milky Way disk. Specifically, as stated in the 
appendix\,C of SFD, the dust colour temperatures
at low Galactic latitudes 
($|b|$ $<$ 5$^{\circ}$) are not well defined and contaminating sources have not been 
fully removed. Furthermore, as pointed out by \citet{Berry2011}, 
most disk stars are embedded in the dust layer, rather than behind it.
As a consequence, the SFD map overestimates the  extinction, making it
unreliable for most disk stars.
Most recently, the $Planck$ team publish an all-sky model of the thermal dust emission
\citep{Abergel2013}, while the Wide-Field Infrared Survey Explorer (WISE) 
collaboration publishes a 
full-sky atlas of the Galactic 12--$\mu$m dust emission \citep{Meisner2013}.

Based on the wide-angle photometric sky surveys in the near-infrared (IR), 
e.g. the  Two Micron All Sky Survey 
(2MASS, \citealt{Skrutskie1997}), a number of studies have 
been carried out to trace the extinction using 
the near-IR colour excess method \citep{Lombardi2001a,Lombardi2011b,
Majewski2011,Rowles2009}. 
 \citet{Froebrich2007} present an integrated extinction map of 127 
$\times$ 63\,deg$^2$
 in the direction of the GAC, with a resolution of 4\,arcmin,
 based on the 2MASS $(J - H)$ 
and $(H - K_{\rm s})$ colours.
Other extinction maps are available for patchy areas 
of the GAC, such as that obtained using 
special tracers from the SDSS 
photometry (e.g. \citealt{Schlafly2010,Peek2010}) 
or spectroscopic data (e.g.
\citealt{Schlafly2011,Jones2011}). However, for 
the purpose of detailed studies of the 
GAC area, a high resolution and homogeneous extinction map covering the 
whole area of interest is highly desired. 

All the extinction maps described above are restricted to
two dimensions and refer to integrated extinction along the line of sight.
As pointed out above, disk stars are embedded in the dust layer. Thus even for 
stars in the same direction, the amounts of extinction they suffer from 
 could be quite different 
due to their different distance information. 
Several approaches are available to obtain information 
of distance and derive an extinction
map in three dimensions (3D).
These include theoretical
dust modeling \citep{Chen1998,Drimmel2003}, 
comparison of near-IR data of giants with stellar population synthesis models
\citep{Marshall2006,Chen2013,Schultheis2013}, ``pair method" based on 
spectroscopic data \citep{Yuan2013b}, and tracing stars 
of special characteristics. 
\citet{Neckel1980} obtain extinction values and distances using UBV, MK and $\beta$
photometric data  for more than 11,000 O to F stars. \citet{Sale2009} and \citet{Sale2012} 
propose a technique to obtain 3D 
extinction maps based on hierarchical Bayesian models.
\cite{Bailer-Jones2011}, \citet{Hanson2013} and \citet{Green2014}
 introduce a similar Bayesian approach for estimating the 
intrinsic stellar parameters and line-of-sight extinction values. 
For a sample of 73 million stars 
with the SDSS photometry 
and a sub-sample of 
23 million stars also with complementary 2MASS photometry,
\citet{Berry2011} estimate distances and values of extinction of individual stars by 
fitting the observed optical (and IR) spectral energy distribution (SED).
Most recently, \citet{Lallement2013} assemble a data set of colour excess
with associated parallax or photometric distances for a sample of 23,000 
nearby stars and construct 
a 3D extinction map of local dust within 2.5\,kpc. 
In the work of \citet{Berry2011}, values of 
extinction up to $A_r$ $\lesssim$ 2--3 
($A_r=0.88A_V$, assuming the extinction law of \citet{Cardelli1989} 
for a total-to-selective extinction ratio $R_V \equiv A_V / E(B - V) = A_V / (A_B - A_V) = 3.1)$
are traced out to a distance of about 2\,kpc. When using only the optical SDSS data,
distances up to a factor of two can be reached using blue main-sequence
stars. For red stars, the distance limit is much smaller. Using only the 2MASS data,
\citet{Froebrich2007} trace $A_V$ up to about 10\,mag for the GAC region. For the
Galactic centre and other high extinction regions, values of $A_V$ can go
 up to at least 20\,mag. \citep{Lombardi2001a,Nidever2012}. 
Studies based on the optical data such as the SDSS are limited to 
$A_V$ of $\lesssim 10$\,mag \citep{Schlafly2010, Peek2010}.

In this paper,  we estimate values of 
extinction and distance for stars in the XSTPS-GAC photometric catalog. We 
apply a method similar to that proposed by \citet{Berry2011} to a sample of 30
million stars with complementary IR photometry from the 2MASS and WISE.
We also estimate extinction using the RJCE method \citep{Majewski2011} based on the 
2MASS and WISE $(H-W2)$ colour. Together with the 
photometric distances deduced from the dereddened XSTPS-GAC optical 
photometry \citep{Ivezic2007,Juric2008}, we
have obtained a 3D extinction map of the 
GAC region with a spatial resolution, depending on latitude, 
that varies  between 3 -- 9\,arcmin.

The paper is structured as following: In Section\,2 we present the relevant 
XSTPS-GAC, 2MASS and WISE data. Section\,3 describes in detail the method 
used to derive values of 
extinction and distance. The results are tested using Monte-Carlo
simulations as well as Galaxy models.
In Section\,4, we present our main results which are discussed in Section\,5.
We summarize in Section\,6. 

\section{Data}

We first describe the data used in the current work.
\citet{Bailer-Jones2011} shows that there is  
a significant degeneracy between the effective 
temperature (or intrinsic colours) and extinction 
derived using just four colours from the artificially $BVJHK$
data for artificially reddened $Hipparcos$ stars. 
\citet{Berry2011} show that the degeneracy can be broken when the $(i-z)$
colour is also available together with $(g-r)$
 and $(r-i)$, assuming a certain extinction law. 
However when only the optical $g$, $r$ and $i$ bands 
are available, such as in the case of 
XSTPS-GAC, the degeneracy can not be broken. 
To overcome the problem, we
introduce a sub-sample of the XSTPS-GAC with matching
IR photometry from the 2MASS and WISE 
to break the
degeneracy. We start by briefly describing the XSTPS-GAC, 2MASS and WISE surveys.

\subsection{The XSTPS-GAC survey}

The Xuyi Schmidt Telescope Photometric Survey of the Galactic Anticentre (XSTPS-GAC),
which started collecting data in October 2009 and completed in March 2011,
was carried out in order to provide input targets for the LSS-GAC \citep{Liu2013}.
The survey collected in
 the SDSS $g$, $r$ and $i$ bands using the Xuyi 1.04/1.20\,m Schmidt Telescope
located on a small hill about 35 km from Xuyi town, 
north of Nanjing, in the middle-eastern area
of China with an elevation of about 180\,m above the sea level. 
The XSTPS-GAC has two components, the main of
5, 400\,$\rm deg^2$ centered on the GAC
covering 3\,h $\le$ R.A. $\le$ 9\,h, and $-10\degr 
\le \rm{Dec.} \le +60\degr$, 
and an extension of about 900 $\rm deg^2$ covering the M 31/M 33 area.
The total survey area is close to 7,000 $\rm deg^2$ including the bridging
fields connecting the two components.
 In total, the XSTPS-GAC archives approximately 100 million stars down to a
limiting magnitude of about 19 in the $r$ band 
($\sim$ 10$\sigma$) with an astrometric accuracy about 0.1$''$
 and a global photometric accuracy of about 2\% \citep{Liu2013}.
The optical broadband atmospheric extinction coefficients
and the night sky brightness of the Xuyi Observational Station of 
the Purple Mountain Observatory  are presented by \citet{Zhang2013}.
 A detailed description of the astrometric calibration of 
the XSTPS-GAC data is given in \citet{Zhang2014}.
In the current work only data 
covering the GAC and from the bridging fields, totaling about 6,000 $\rm deg^2$,
are used. A
full account of the observation and data processing will be presented
in a future contribution.

The observations were carried out with a thinned 1,096 $\times$
 4,096 CCD camera.
The effective field-of-view (FoV) of the camera is
1.94$^{\circ}$ $\times$ 1.94$^{\circ}$ with a sampling of 1.705$''$ 
per pixel projected on the sky.
The integration time was 90\,s. The readout time was 43\,s using 
the slow dual channel readout mode.
The adjacent fields stepped by $0.95\degr$ in RA (i.e. yielding 50\% 
overlap) and $1.90\degr$ in Dec.
Typically two adjacent stripes along the RA were scanned field by field 
alternatively. With this observing strategy,
the hour angle and zenith distance of the telescope were nearly 
constant as the observation processed,
yielding maximum uniformity. To facilitate the ubercal (global flux 
calibration), a number of ``Z-stripes'' were added.
The ``Z-stripe'' fields straddled between two ``normal'' stripes (i.e. stripes along the RA),
to help tie the latter together on a common flux scale.
All observations were carried out under dark/grey lunar conditions,
with air masses lower than 1.3, except for fields of very low declinations.

A preliminary astrometric calibration 
was first carried out with the 
 Guide Star Catalog\,2.0 (GSC\,2.0; \citealt{Morrison2001})
as the reference using 
an eight-parameter linear model,
yielding an accuracy of about 0.5$''$. The plate distortion was then 
modeled with a 30-parameters two-dimensional function
using the PPMXL as the reference catalog, reaching an accuracy of about 
140\,mas for individual frames.
By combining all frames, we expect to achieve a final astrometric accuracy 
of about 60\,mas. 
The frames were flat-fielded using super sky flats generated from target frames 
by clipping all individual stars.
Aperture and Point Spread Function (PSF)
 photometry were carried with a DAOPHOT-based pipeline. 
A ubercal flux calibration against
the SDSS photometry using overlapping fields a photometric accuracy of better than 2\% 
 for a single frame and about 2 -- 3\%
for the whole survey sky area.
The catalog generated by XSTPS-GAC have been used 
to select targets during  the LAMOST 
commissioning phase as well
as for the Pilot and the on-going Regular Survey of LSS-GAC.
The data will also be used in the reduction and follow-up analyses of the LAMOST data.
In addition, taking advantage of its coverage,
depth and photometric accuracy,
the data will be an asset to study the disk structures
(e.g., scale heights and lengths of the thin and thick disks; disk flares, warps
 and truncation)
and constrain the sub-structures (e.g., the Monoceros 
Ring) in the GAC direction.

\begin{figure*} 
\centering
   \includegraphics[width=\textwidth]{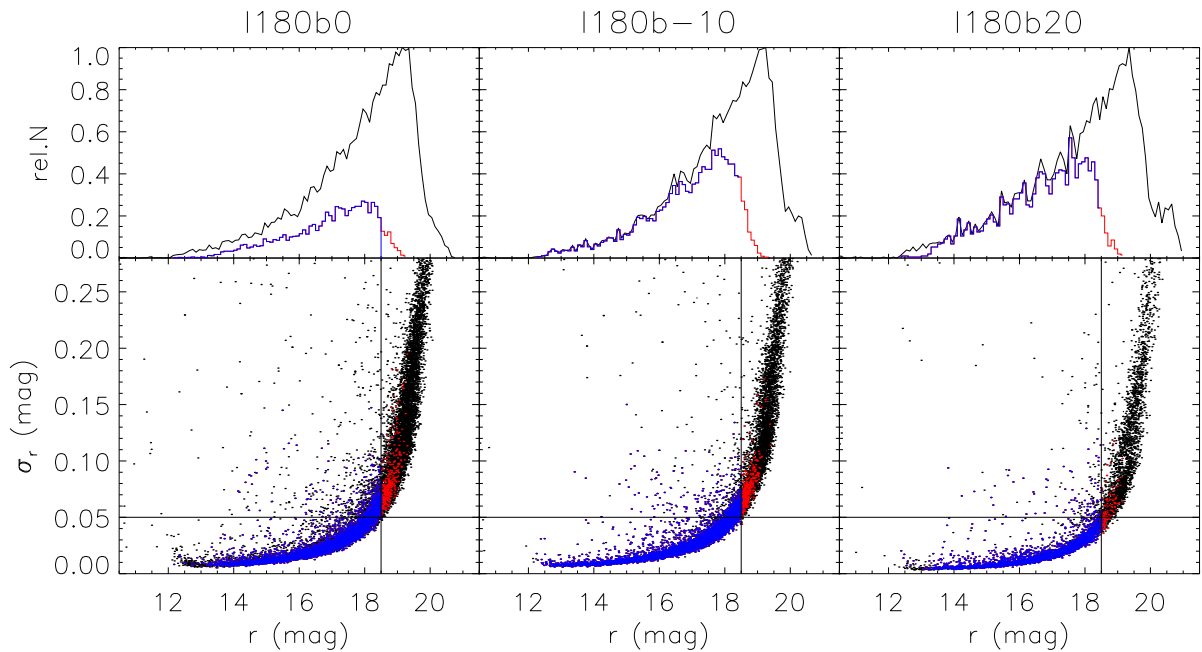}
\caption{
Distributions of relative star count (upper panels)
 and (formal) $r$-band photometric measurement uncertainties (
lower panels) as a function of $r$-band magnitude for three fields, 
each of $1^{\circ}\times 1^{\circ}$, for lines of sight 
$l = 180^\circ$ and $b = 0$, $-10$ and $+20^{\circ}$, 
respectively. The black lines and dots represent all 
XSTPS-GAC sources in the fields, whereas those represented by 
blue plus red lines and dots are sources with matching 2MASS and WISE 
photometry. Only sources represented by blue lines and dots are included 
in the final sample analyzed in the current work.}
\label{fig1}

\end{figure*}

In the bottom three panels of
Fig.\,\ref{fig1} we plot the photometric error and magnitude relation for $r$ band 
for three different lines of sight ($l=180^{\circ}$, $b=0$, $-10$, 20$^{\circ}$) as
examples. Black dots represent XSTPS-GAC measurements. Typically as
$r$ band magnitude reaches 18.5, the photometric errors rise to 0.05\,mag
(see the black lines in the three panels).
 The black lines in the upper three panels of Fig.\,\ref{fig1} plot
the normalized star counts as a function of $r$
magnitude, for the XSTPS-GAC data. 
The distributions show the survey is complete down to a limiting magnitude fainter 
than 19\,mag in $r$-band. 
For sources of $r$ $<$ 16.5\,mag, the 
measurement errors become smaller than 0.02\,mag, 
which is the estimated systematic error of the global flux calibration of XSTPS-GAC. 
To account for the calibration uncertainties, all errors are reset to 
0.02\,mag when the reported errors are smaller
than 0.02.

\subsection{The 2MASS Survey}

The Two Micron All Sky Survey (2MASS)
 surveyed the entire sky using two 1.3-m  aperture telescopes at Mt. Hopkins and CTIO, 
Chile \citep{Skrutskie1997}.  
Each 2MASS camera contained three NICMOS3 256 $\times$ 256 HgCdTe detectors,
collecting images in three near-IR bands simultaneously: 
$J$, $H$ and $K_{\rm s}$ bands centered at 
1.25, 1.65 and 2.16\,$\rm \mu$m, respectively. For point-sources,
a signal to noise ratio  S/N = 10 limit corresponds
 to a limiting magnitude at $J$ = 15.8, $H$ = 15.1, 
and $K_{\rm s}$ = 14.3\,mag for virtually the entire sky.
Sources brighter than the above limiting magnitudes in the 2MASS Point Source Catalog
are believed to be highly complete ($>$0.99) and reliable ($>$0.9995). 
The photometric systematic uncertainties of  2MASS are estimated to
 be smaller  than 0.03\,mag, and astrometric 
uncertainty for these sources is less than 0.2$''$.

\subsection{The WISE survey}

The Wide-field Infrared Survey Explorer (WISE) surveys the entire sky
with a 40\,cm telescope on board  
the satellite \citep{Wright2010}. 
WISE maps the whole sky in four infrared bands 
$W_1$, $W_2$, $W_3$ and $W_4$ centered at 3.4, 4.6, 12 
and 22\,$\mu$m respectively, with corresponding spatial 
angular resolution of 6.1, 6.4,
 6.5, and 12.0$^{\prime\prime}$ in the four bands. 
The WISE All-Sky Data Release area is comprised of 18,240 Atlas Tiles, 
with each Tile spanning 1.564$^\circ$ $\times$ 1.564$^\circ$ in 4095 $\times$ 4095 pixels 
at a resolution of 1.375$''$ per pixel.
The WISE Source Catalog contains positions and photometry for about 564 million point-like 
and resolved objects. All sources cataloged have a measured S/N greater than
5 in at least one of the four bands.
The positions are calibrated against the 2MASS, achieving 
an accuracy of $\sim$200\,mas on each axis with respect to the 2MASS 
reference frame for sources with S/N better than 40. 
The 5$\sigma$ photometric sensitivities are 0.068, 0.098, 
0.86 and 5.4\,mJy (equivalent to Vega magnitudes 16.6, 15.6, 11.3 and 8.0, respectively) at 
3.4, 4.6, 12 and 22\,$\mu$m, respectively, in unconfused regions in the ecliptic 
plane. Saturation affects photometry of sources brighter than approximately 
8.1, 6.7, 3.8 and $-0.4$\,mag at 3.4, 4.6, 12 and 22\,$\mu$m, respectively.   
We use only $W1$ and $W2$ bands in the current work because of 
 the low sensitivities of $W3$ and $W4$ bands and the poor 
angular resolution of the $W4$ band.

\subsection{Sample selection}

We select sources in the XSTPS-GAC
with matching targets in the 2MASS and WISE catalogs. 
We use the Centre de Donnes astronomiques de Strasbourg (CDS) XMatch 
Service\footnote{http://cdsxmatch.u-strasbg.fr/xmatch} to 
cross-match the sources by position on the sky. 
The matching radius is set to 1.5$''$, with which 
the fraction of multiple matches is less than 0.01\% for 
the whole GAC survey area. About 90\%
of the cross-identified sources have a matching distance between 
the XSTPS-GAC and 2MASS/WISE positions smaller than 0.5$''$.
The combined XSTPS-GAC/2MASS/WISE catalog contains about 30 million sources. 
As mentioned before, XSTPS-GAC sources with $r$ $<$ 18.5\,mag typically have photometric
errors smaller than 0.05\,mag (c.f. Fig.\,\ref{fig1}). To select sample stars from the 
combined XSTPS-GAC/2MASS/WISE catalog,
we require that the sources must have $r$ band magnitude 
$r$ $<$ 18.5\,mag and are detected in all bands, i.e. XSTPS-GAC $g$, $r$, $i$; 
2MASS $J$, $H$, $K_{\rm s}$; and WISE $W1$, $W2$.
The requirements lead to approximately 27 million sources ($\sim$90\%) 
 in the combined catalog.

For three selected lines of sight, the distributions 
of $r$-band star count as well as (formal) photometric measurement uncertainties
of XSTPS-GAC sources, 
those with matching 2MASS and WISE photometry 
(blue plus red lines and dots), and those included 
in the final sample (blue lines and dots only) are illustrated
in Fig.\,\ref{fig1} as examples.
As the Galactic latitude decreases, the number of star counts increases, yet
the fraction of XSTPS-GAC with matching
2MASS and WISE photometry
becomes smaller. At the Galactic plane ($b=0^\circ$, left panel of Fig.\,\ref{fig1}),
only $\sim$ 45\% of XSTPS-GAC sources have near- and mid-IR magnitudes.   
The fraction also decreases for fainter sources.
There is a steep 
decrease around $r$ = 18.5\,mag. 
This is not surprising given that
the survey depth of XSTPS-GAC is deeper than 
those of 2MASS and WISE,
at least for not so heavily reddened regions. 
The XSTPS-GAC/2MASS/WISE combined catalog has a typical completeness
limiting $r$-magnitude slightly deeper  
than 18\,mag. Main-sequence stars with 2MASS $K_{\rm s}$ $<$ 14.3 
have distances smaller 
than approximately 1--2\,kpc \citep{Berry2011}, while dwarfs with XSTPS-GAC
$r$ $<$ 18.5\,mag can be as far as 4\,kpc \citep{Liu2013}.  
As in the case of  the three optical bands of XSTPS-GAC, for the 
three 2MASS near-IR and two WISE mid-IR bands, 
the errors of sources with (formal) reported photometric errors
less than 0.02\,mag are reset to 0.02\,mag. 

\section{Method}

In this work we adopt a similar method originally proposed by \citet{Berry2011},
by applying SED fitting to multi-band data from the XSTPS-GAC, 2MASS 
and WISE combined catalog.
There are two 
empirical results that form the basis of this method: the stellar locus in the 
multi-dimensional colour space and the shape of dust extinction curve
being depicted by a one-parameter function. 
We begin introducing our method by first describing those two empirical 
results. 

\subsection{Reference stellar locus}

The nearly blackbody emission spectra of stars place them 
predominantly along a line in optical and infrared colour–-colour space.
The broad band colours of a star are almost entirely determined by the 
star's effective temperature, metallicity, and surface gravity. 
Furthermore, the optical and infrared colours are
largely determined by the effective
temperature and hardly affected by the surface 
gravity and metallicity \citep{Covey2007,Ivezic2007,High2009}. 
As a consequence, in the treatment below, we have neglected the 
effects of the latter two parameters. 

\citet{Covey2007} calculate the running-median of colours for a sample of 
600,000 stars, observed by the SDSS and 2MASS in the 
$ugrizJHK_{\rm s}$ photometric system. 
The extinction of sample stars are
 estimated using the
SFD maps for $A_r$ $<$ 0.2\,mag. They present the
stellar locus as a function of colour $(g-i)$,
for the range $-0.25~<~(g-i)~<4.50$. 
The results are used by \citet{Berry2011} as an empirical SED library and adopted by
\citet{High2009} as the standard empirical stellar locus. 
Most recently, \citet{Davenport2014} 
present the fiducial main sequence stellar locus traced by 
10 photometric colors observed by SDSS, 2MASS, and WISE. 
Because the photometric catalog of  XSTPS-GAC
 is an entirely new data set and \citet{Covey2007} lack the information of mid-IR 
colours, we have recalculated
 the empirical stellar locus for our own XSTPS-GAC, 2MASS  and WISE
combined multi-band catalog. We first define a
``high-quality reference sample of (essentially) zero extinction'' 
by imposing the following criteria:
a line-of-sight Galactic extinction $A_r$ less than 0.075\,mag 
as given by the SFD map and photometric
errors  in the individual band ($g$, $r$, $i$, $J$, $H$, $K_{\rm s}$, 
$W1$, $W2$) smaller than 0.05. 
The criteria are more restrictive than those of \citet{Covey2007} and
\citet{Davenport2014}. The cuts lead to
a reference sample consisting of 132,316 stars.
As in \citet{Covey2007}, we assume the $(g-i)$ colour as the 
independent variable for the stellar locus. 
Given the above strict cut on the upper limit of extinction, 
we have ignored the 
effect of reddening for the reference sample.

\begin{figure*}
\centering
   \includegraphics[width=0.80\textwidth]{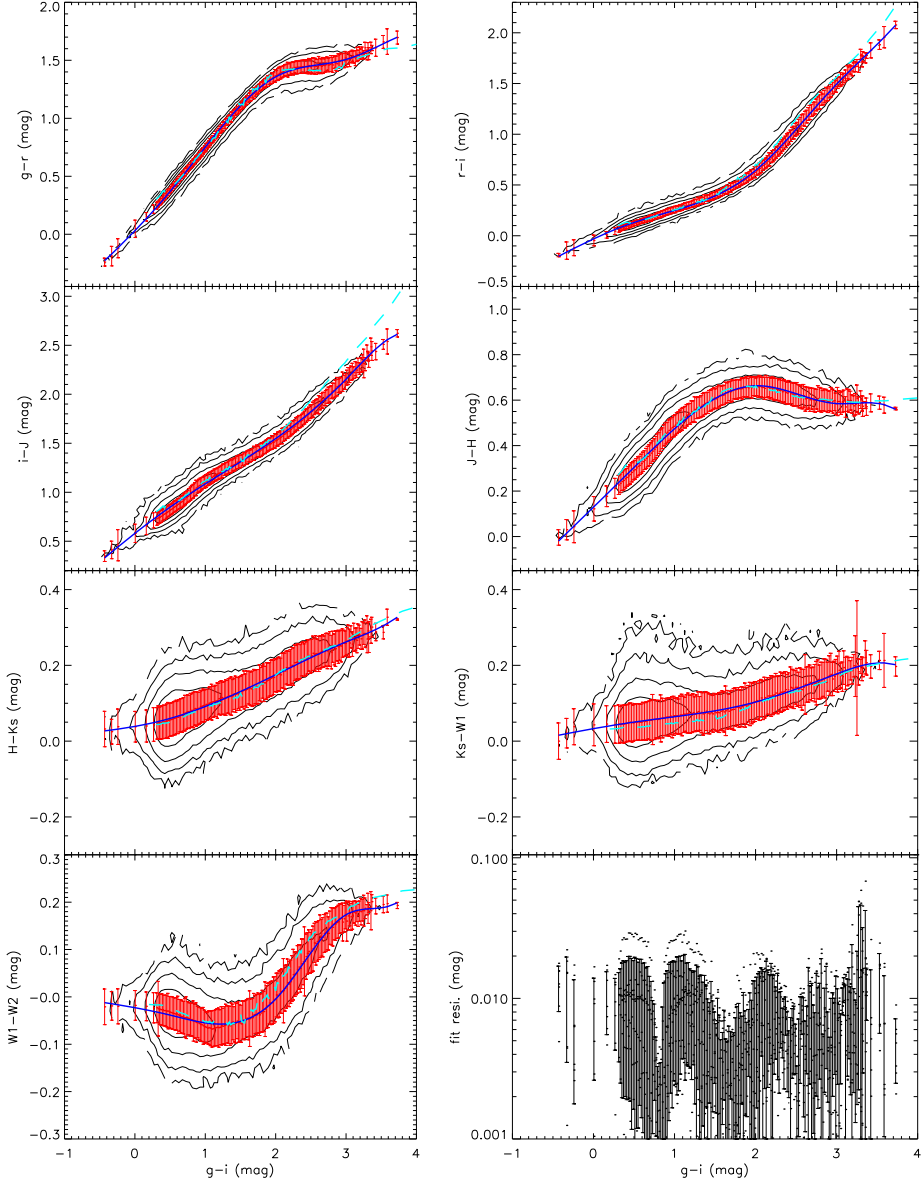}
\caption{Colour-colour diagrams of 132,316 reference 
stars of essentially zero interstellar extinction selected 
from the combined catalog of the XSTPS-GAC, 2MASS and
 WISE surveys with high quality photometry. The contours are on a logarithmic scale.
Red pluses and associated errorbars represent the median values and standard 
deviations of the colour plotted as ordinate for the individual bins of $(g - i)$ as abscissa.
 The blue lines are spline fits to the
median values. Also overplotted for comparison are stellar locus from
\citet{Davenport2014}. The bottom-right
panel shows the residuals of spline fits with the 
errorbars indicating the scatters of residuals of the seven 
colours.}
\label{fig2}

\end{figure*}

The colour-colour
diagrams for our reference sample are plotted as density contours in Fig.\,\ref{fig2}.
The abscissa is the chosen independent variable $(g-i)$ while the ordinates include 
the seven colours from adjacent bands, $(g-r)$, $(r-i)$, $(i-J)$, $(J-H)$, $(H-K_{\rm s})$, 
$(K_{\rm s}-W1)$, and $(W1-W2)$. The contours are shown
on a logarithmic scale.
Fig.\,\ref{fig2} shows that the stellar locus are very compact.
The locus for the IR bands, $(H-K_{\rm s})$, $(K_{\rm s}-W1)$, and $(W1-W2)$ 
appear to be slightly broader. This is 
because of the small colour ranges of the ordinate. 
We notice that our multi-band catalog, $(g-i)$ spans a range from about $-0.5$ to 4. 
We then calculate the median values for each colour of adjacent bands 
in individual $(g-i)$ bins 
ranging from $-0.5$ to 3.8. For $0.2~<~(g-i)~<~3.48$, the bin 
width is 0.02\,mag, and 0.1\,mag
outside the range considering
the low stellar density. The
median values are represented by red crosses in Fig.\,\ref{fig2}. 
Overplotted in blue are spline fits 
to the median values. Even a 5th order polynomial can not fit
the data well, 
in particular near the blue and red edges of the stellar
locus. The adopted spline function fits the data
very well, with typical residuals $\sim 
0.01$\,mag and less than 0.02\,mag in general (c.f. the right upper panel of Fig.\,\ref{fig2}).
The resultant fits are used to generate a standard SED library
 adopted in the current work. 
For comparison, we also plot in Fig.\,\ref{fig2}, 
 the locus obtained by Davenport et al. (2014;
cyan dashed lines) colours of the three optical and three near-IR bands. 
The agreement is very good, although some small differences 
are clearly visible towards redder colors. 

Except for $(J - H)$ and $(W1 - W2)$, one of the 
most significant features of the standard locus
plotted in Fig.\,\ref{fig2} are their monotonic behavior. 
Some locus, including those of $(g-r)$, $(r-i)$, $(J-H)$ and $(W1-W2)$
 exhibit a change of slope at around $(g-i)$ $\sim$ 1.95\,mag.
As shown by previous studies, 
\citep{Finlator2000, Hawley2002, Covey2007,
 Juric2008, High2009}, blueward of this break point, 
evolved and main-sequence A- to K-type stars dominate 
the population, whereas redward of the point M-type dwarfs take over.
The metallicity dependence of the disk main-sequence stars
blueward of this break point of locus is small. The metallicity variations 
of M\,dwarfs in the
disk could perturb the kink region of the observed locus \citep{High2009}. 

The standard SED library
generated from the reference stellar locus
 can not be applied for  blended or binaries, which may possess quite different 
SEDs \citep{Richards2001, Smolcic2004,
Eisenstein2006, Berry2011}.
Fortunately as already discussed by \citet{Berry2011}, those stars 
consist only a minority
\citep{Finlator2000, Juric2008, Liu2013} 
and can generally be detected and excluded by their 
large values of minimum $\chi^2$ in the process of SED fitting.

\subsection{Extinction law}

\citet{Cardelli1989} derive a $R(V)$-dependent Galactic extinction law for
the wavelength range $0.125~ \leq ~ \lambda ~ \leq ~ 3.5\,\mu$m.
Slightly different results are obtained by \citet{ODonnell1994} for 
the near-UV to the optical wavelength range ($0.303~ \leq~ \lambda~ \leq ~ 0.909\,\mu$m). 
A new extinction law from the UV to the IR 
($0.1 \leq \lambda \leq 3.4\,\mu{\rm m}$) is presented by 
\citet{Fitzpatrick1999}. Following the releases of survey data of
the SDSS in the optical and other surveys in the 
IR, such as 2MASS in the near-IR and GLIMPSE 
(The Spitzer Galactic Legacy Infrared Midplane Survey Extraordinaire;
 \citealt{Benjamin2005}) in the mid-IR, number of studies have been
carried out investigating the extinction law in the optical and 
IR wavelength ranges for different regions and environments of the Milky Way
 (e.g. \citealt{ Indebetouw2005, Jiang2006, Flaherty2007,
Nishiyama2009, Fitzpatrick2009, Gao2009, Schlafly2011, Chen2013}). 
Using SDSS and 2MASS data, \citet{Berry2011} 
constrain the shape of the extinction curve  and find that 
their result is compatible to that of \citet{Fitzpatrick1999} 
but not of \citet{ODonnell1994}. 
\citet{Berry2011} recommend the fixed $R_V$ value as 3.0 and
 a corresponding interstellar extinction 
coefficient $C_\lambda \equiv A_\lambda/A_r$ of 
1.400, 0.759, 0.317, 0.200 and 0.132 for $g$, $i$, $J$, $H$, $K_{\rm s}$
bands, respectively.    

More recently, \citet{Yuan2013} 
combine data from the SDSS, GALEX (Galaxy Evolution 
Explorer; \citealt{Martin2005}), 
2MASS and WISE, and study
the extinction law of the 
Milky Way from the far UV to the mid-IR using the star pair technique. 
The technique pairs stars
suffering from high extinction 
with their twins that suffer from almost nil extinction 
but otherwise
have almost identical stellar parameters
as given by spectroscopic observations and analyses. 
Using this technique, Yuan et al. measure the dust extinction
for thousands of Galactic stars and derive the
empirical reddening coefficients for 
photometric bands ranging from the UV to the mid-IR.
The method has the advantages that it is straight-forward,
model-free and applicable to the majority of stars.
For the optical and IR bands, 
they obtain values of $C_{\lambda}$ 
= 1.42, 0.74, 0.31, 0.19, 0.13, 0.082 and 0.065
for $g$, $r$, $J$, $H$, $K_{\rm s}$, $W1$ and $W2$ bands,
respectively, very similar to those of \citet{Berry2011}.
Considering that the current work use the same optical and IR data set
as \citet{Yuan2013}  and the fact that 
the stellar locus obtained here are almost identical to those derived using data 
from the SDSS \citep{Covey2007, Davenport2014}, we have therefore simply adopted 
the extinction coefficients given by \citet{Yuan2013} of a fixed $R_V$ value of 3.1.

\subsection{Extinction determinations}

Our procedure to derive extinction by SED fitting
 is very similar to that of \citet{Berry2011}.
With the reference  stellar locus and extinction law fixed, 
assuming a SED library index $t$ 
[in our case, we use $(g-i)_0$] and a $r$-band
extinction $A_r$, we can apply a simple model to simulate the 
observed colour of bands $\lambda 1$ and $\lambda 2$
 for a given star \citep{Berry2011}: 
\begin{equation}
c_{sim}=c_{sl}[(g-i)_0]+(C_{\lambda 1}-C_{\lambda 2})A_r
\end{equation}
where $c_{sim}$ is the simulated colour, $(g-i)_0$ 
describes the position of the star 
on the reference stellar locus, $c_{sl}$ is
the colour predicted by the reference 
stellar locus for $(g - i)_0$, 
$C_{\lambda 1}$ and $C_{\lambda 2}$ are  
extinction coefficients for bands $\lambda 1$ and $\lambda 2$ given by
the assumed extinction law. With 
$(g-i)_0$ and $A_r$ as free parameters, we can 
then model the seven
observed colours formed by adjacent bands
(e.g. $(g-r)$, $(r-i)$, etc.) of our data set. Similar to  Eq.\,(3) 
of \citet{Berry2011}, we define 
\begin{equation}
\chi^{2} = \frac{1}{5} \sum_{i=1}^7 (\frac{c_{obs} ^i - c_{sim} ^i}{\sigma_i})^2
\end{equation} 
where $c_{obs} ^i$ and $c_{sim} ^i$ are
the $i$th observed and simulated
colours ($i$=1, 2, ..., 7), and $\sigma _i$ is 
the colour measurement uncertainty 
computed by adding the photometric errors
in quadrature. The best-fit model 
is then obtained by minimizing $\chi ^2$.

The optimization to minimize $\chi ^2$, 
is carried out  by running a 
pseudo $A_r$ ranging from 0 to 10\,in step of 0.02\,mag and 
all index value $i~ \equiv ~ (g-i)_0$
of our reference SED library. 
The procedure yields best fit values of $A_r$, 
as well as $(g-i)_0$ for the star of concern.
In doing so we have ignored the possible contributions 
to colour uncertainty $\sigma_i$ produced  by the
finite width of the empirical library SED 
, a simplification also made by 
\citet{Berry2011}. The errors of the
 best fit  values of $A_r$ and $(g-i)_0$ are calculated from the 
ellipse fitting parameters of the $\chi ^2$ surface defined by 
$\chi ^2 < \chi_{\rm min}^2 + 6.17$, 
see \citet{Berry2011} for more information.

\subsection{Extinction and photometric parallax of open clusters}

Distances of stars in our sample are calculated using the derived 
best-fit values $(g-i)_0$ using a  
photometric parallax relation, such as that given by \citet{Juric2008} and \citet{Ivezic2008}.
This is a simplified approach.
In addition to ignoring possible 
binarity and blending, the approach also ignores the fact that stars
have different metallicities and ages, and assumes that 
 all stars are main-sequence dwarfs. Accordingly, the
best-fit parameters, $(g-i)_0$ and $A_r$ returned by the procedure could occasionally be 
totally wrong. We have made a series of 
tests to check the reliability of our method. As a first test we apply the method to
a few well studied stellar clusters.

We select two open clusters in the GAC region: M\,67 (NGC\,2682) at
high Galactic latitude ($l$ = 215.696\,deg, $b$ = +31.896\,deg) with 
low dust extinction \citep{Xin2005} and 
the other is M\,35 (NGC\,2168) at low Galactic latitude ($l$ = 186.591\,deg, $b$ = +02.191\,deg,) 
with high dust extinction  \citep{Wu2009}. Basic parameters of those
two clusters, shown in Table\,\ref{ta1}, are well documented in the literature, such as the 
recently published papers by \citet{Kalirai2003, Yadav2008, An2008, 
Geller2010}. 

\begin{table*}
 \centering 
  \caption{Basic parameters of M\,67 and M\,35.}%}
  \begin{tabular}{cccccc}
  \hline 
  \hline
 & distance (pc) & $(m-M)_0$ (mag) & age  & $\rm{[Fe/H]}$ & $E(B-V)$ \\
 \hline 
M\,67 & 890 $^{a}$ & 9.56 -- 9.72 $^{b}$ & 3.5 -- 4.8\,Gyr $^{b}$ & 
$+0.03 \pm 0.01$ $^{c}$ & 0.041$^{d}$ \\
M\,35 & $\rm 912_{-65} ^{+70}$ $^{e}$ & $9.80 \pm 0.16$ $^{e}$ & 180\,Myr $^{e}$ 
& $-0.21$ $^{e}$ & 0.20 $^{e}$ \\
 \hline
\end{tabular}\\
Reference: $a$: \citet{An2008}; $b$: \citet{Yadav2008}; 
$c$: \citet{Randich2006}; $d$: \citet{Taylor2008}; $e$: \citet{Kalirai2003}.
  \label{ta1}
\end{table*}

\begin{figure*}
\centering
   \includegraphics[width=15cm]{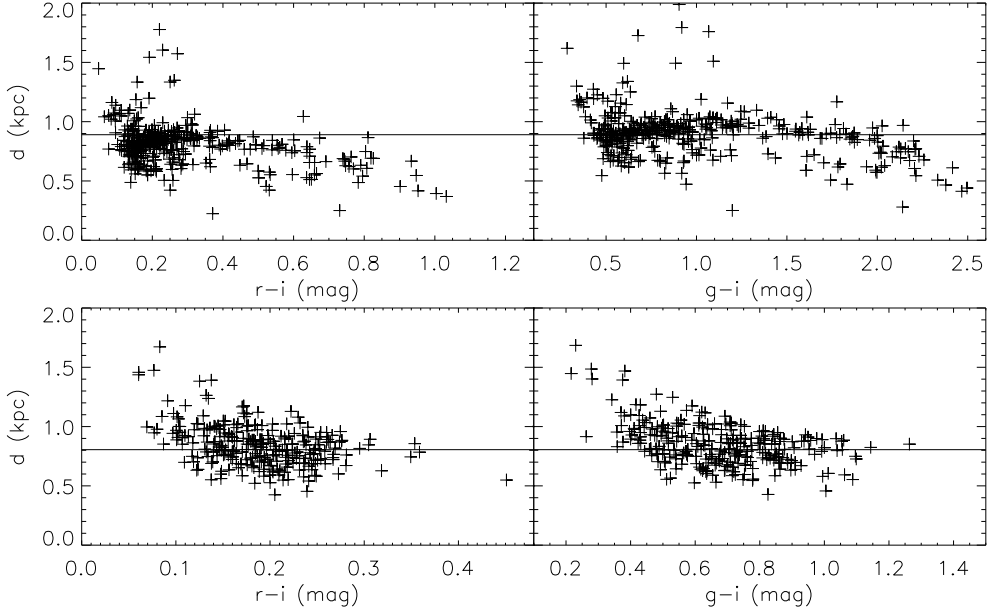}
\caption{Distances deduced for member stars of M\,35 (bottom panels) 
and M\,67 (upper panels), using the ``bright'' relation of 
Juri\'{c} et al. (2008; left panels) and that of 
Ivezi\'{c} et al. (2008; right panels). The horizontal 
lines denote reference distances adopted for the two clusters, 
0.89\,kpc for M\,67 and 0.91\,kpc for M\,35.} 
\label{mcdist}

\end{figure*}

The sources of M\,67 are selected using the cross-match between our sample and
 the catalog of \citet{Yadav2008}.
We use sources with a proper-motion membership probability
larger than 60\% in their catalog and obtain 403 stars. Members of M\,35
are selected by matching with the catalog of \citet{Geller2010}. We select
stars with a radial-velocity membership probability
larger than 50\% and obtain 282
stars. We then apply our method to the selected stars and obtain 
best-fit values of $(g-i)_0$ and $A_r$ for each member star 
in the two clusters. 
Regarding our new stellar locus are in good agreement with those deduced from the
SDSS data \citep{Covey2007,Davenport2014},
distances of individual stars are 
calculated using two photometric parallax relations, one from the 
``bright'' relation of \citet{Juric2008} 
and another from Ivezi\'{c} et al. (2008). 
The ``bright'' relation is given in terms of $(r - i)$, 
which can be converted from the best-fit 
value of $(g - i)_0$ using the reference stellar locus:
\begin{eqnarray*}
M_{r,J08} & = & 3.2+13.30(r-i)-11.50(r-i)^2 \\
          &   & 5.40(r-i)^3 - 0.70(r-i)^4.
\end{eqnarray*}
The relation of \citet{Ivezic2008} is given in
terms of $(g-i)_0$ colour:
\begin{eqnarray*}
M_{r,I08} & = &  \Delta M_r({\rm[Fe/H]}) -5.06+14.32(g-i) \\
          &   &  -12.97(g-i)^2 + 6.127(g-i)^3 - 1.267(g-i)^4 \\
          &   &  +0.0967 (g-i) ^5,
\end{eqnarray*}
where $\Delta M_r({\rm [Fe/H]})~=~4.50-1.11{\rm [Fe/H]}-0.18{\rm [Fe/H]}^2$ 
is the metallicity correction.
The distance $d$, in units of parsec, of a star with
an intrinsic absolute magnitude $M_r$, can
be calculated from
the standard relation, $d=10^{0.2(r_0 - M_r)+1}$, where $r_0=r-A_r$ is the 
dereddened magnitude of the star in $r$ band.
In Fig.\,\ref{mcdist}, we plot distances calculated from the two photometric
parallax relations for member stars of the two clusters.
The black horizontal lines in the diagram denote the 
reference distances adopted for the two clusters: 0.89\,kpc for M\,67 and 0.91\,kpc for
M\,35. Both relations seem to yield satisfactory results for 
members of the two clusters except for a  
few outliers. There is a small systematic offset between the
distances yielded by the two
relations, those deduced using the \citet{Juric2008} relation
($d_{J08}$) are about 0.1\,kpc smaller than values derived 
from the \citet{Ivezic2008} relation ($d_{I08}$). 
Fig.\,\ref{mcdist} also shows there is a linear trend between $d_{J08}$
and $(r-i)_0$ -- redder stars have smaller distances. 
Such a trend is not so obvious for $d_{I08}$. In this study, we will adopt
distances calculated from the best-fit $(g - i)_0$ colour using
the \citet{Ivezic2008} parallax relation,
which is applicable for a wide $(g - i)_0$ colour range from 0.0 to about 4.0 
\citep{Ivezic2008}. For the term of metallicity correction in the \citet{Ivezic2008} 
relation, since almost all stars in our sample are from the disk,
we have adopted a universal ${\rm [Fe/H]} = -0.2$
for all stars, the median metallicity for stars targeted by the 
LSS-GAC (Xiang et al. 2014, in preparation). As shown by Xiang et al. (2014) 
most stars targeted by the 
LSS-GAC have $[{\rm Fe/H}]$ between $-0.7$ and 0.2\,dex.
For a star with an extreme value of metallicity 
($[{\rm Fe/H}] = -0.7$ or $+0.2$\,dex), adopting the 
universal [Fe/H] value of $-0.2$\,dex will 
lead to a metallicity correction term 
$\Delta M_r([{\rm Fe/H}])$ offset by about $\pm$0.45\,mag, and consequently a  
distance offset by about $\pm$0.2\,kpc, or about 20\%
error for a star at 1\,kpc.

\begin{figure}
\centering
   \includegraphics[width=0.48\textwidth]{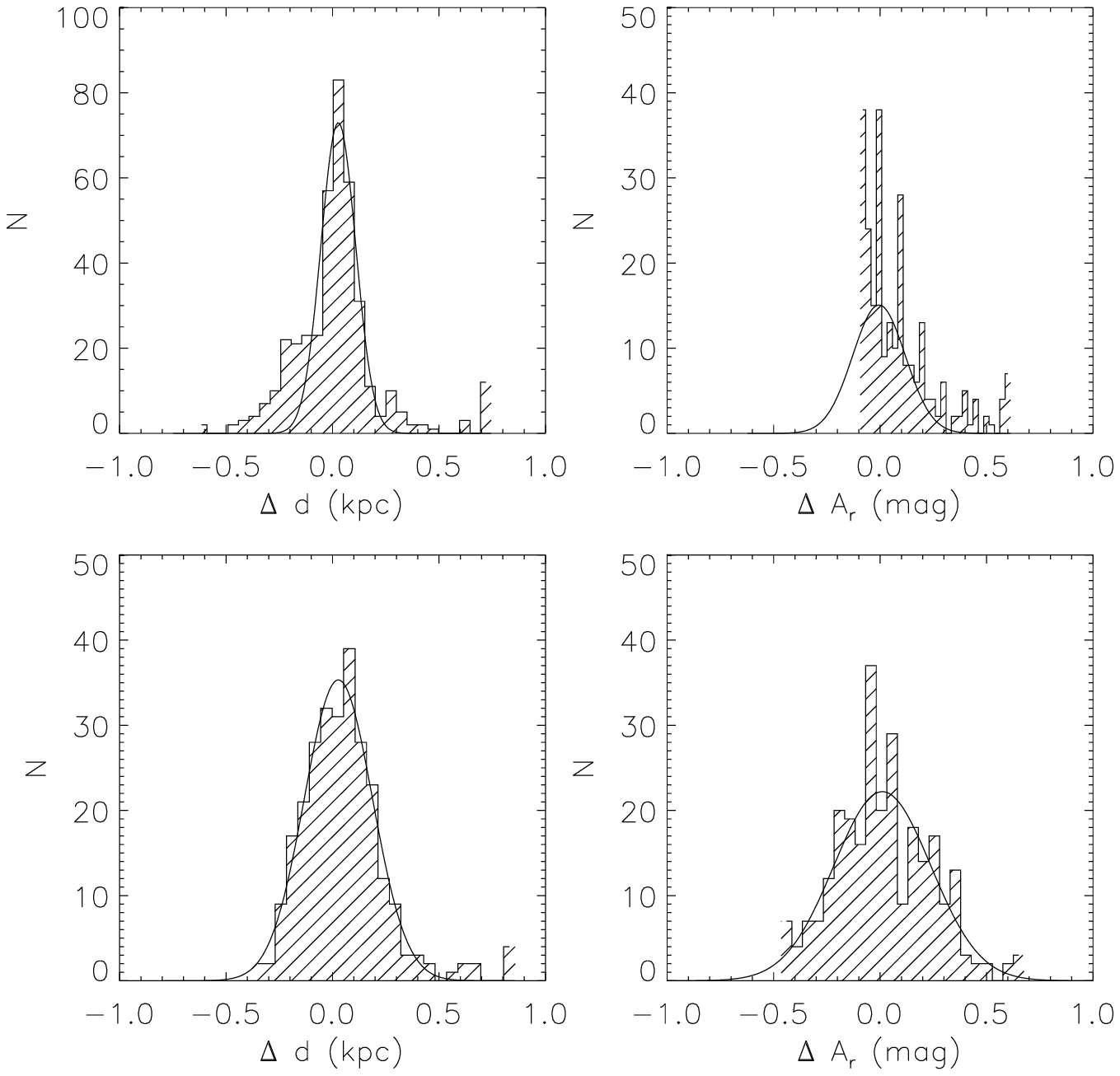}
\caption{Distributions of offsets of values of distance $d$ (left panels) and extinction 
$A_r$ (right panels) relative to 
 the reference values for M\,35 (bottom
panels) and M\,67 (upper panels).
Also overplotted are Gaussian fits to the distributions.}
\label{mcebv}

\end{figure}

Fig.\,\ref{mcebv} shows the distribution of 
distances $d$ and 
extinction $A_r$ 
deduced for individual stars relative to 
the reference values for the two clusters.
Also overplotted are Gaussian fits to the
distributions. The good fits show that
our method is capable of yielding reliable 
parameters for most stars.
The absence of stars on the left of the distribution of 
$\Delta A_r= A_{r,{\rm fit}} - A_{r,{\rm ref}}$ in the case 
 of M\,67 is due to 
the very low extinction of this cluster
($A_{r,\rm ref}~=~0.093$\,mag) and
the fact that negative values of extinction yielded by the 
fitting process are deemed unphysical and therefore discarded.
We also note that there is a systematic offset 
in the distributions of distances, in the sense that distances 
yielded by the photometric parallax relation are on average about
0.07\,kpc larger than the reference values. For extinction, no obvious 
systematic offsets are seen. The broader distributions of M\,35 
compared to those of M\,67 may indicate that 
higher extinction has an impact on both distance and extinction estimates.

\subsection{Test with fiducial stars and mock catalogs}

In this Section, we perform further tests on the reliability and accuracy of
 our algorithm in recovering the reddening and distances 
by applying the method to simulated 
fiducial stars, as well as to 
mock photometric catalogs created with the 
Besan\c{c}on Galaxy model \citep{Robin2003,Robin2012}.

\begin{figure*}
\centering
\includegraphics[width=15cm]{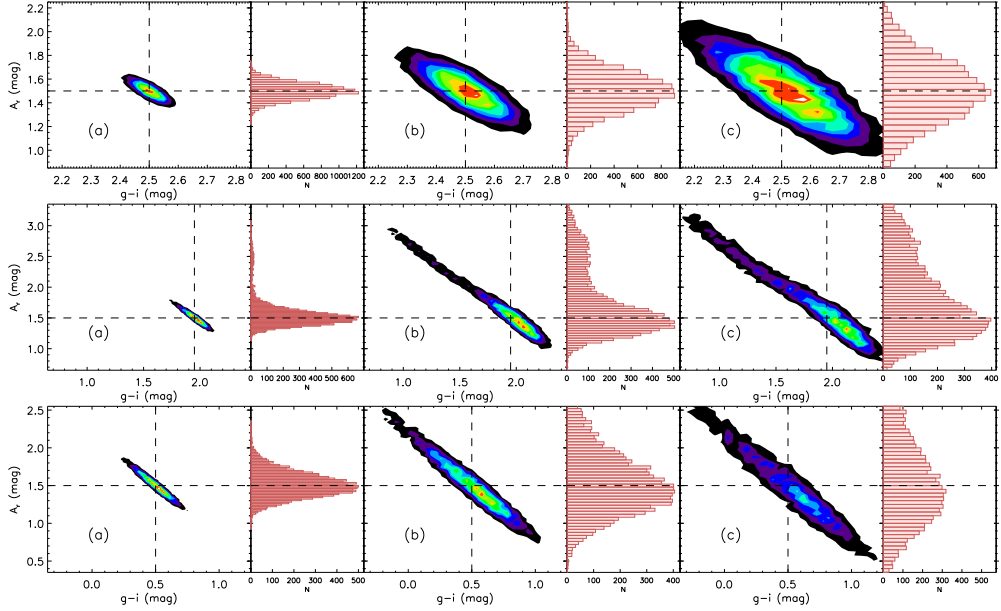}
\caption{Monte Carlo simulations for 
three fiducial stars of intrinsic colours 
$(g - i)_0 = 2.5$ (top panels), 1.95 (middle 
panels), and 0.5 (bottom panels). The pseudo colour 
image in each panel shows  distribution of the best-fit parameters
in the $(g - i)_0$ and $A_r$ plane. Also shown to the 
right of each panel is a histogram of the best-fit values 
of $A_r$. For each row, (a), (b) and (c) from left to right the 
three panels refer respectively to the results 
of the three assumed cases of photometric errors, 0.02, 
0.05 and 0.08\,mag.}
\label{mcfit}

\end{figure*}

First we simulate a fiducial star with different intrinsic colours 
$(g-i)_0~ =~ 0.50,~1.95$ and $2.50$, representing stars located on the 
blue, ``knee'', and red positions of the stellar locus (c.f. Fig.\,\ref{fig2}). 
Three sets of photometric errors
 are assigned using Gaussian distributions of widths 
 0.02, 0.05 and 0.08\,mag, respectively.
The value of 
$A_r$ is fixed at 1.5\,mag. Photometric
magnitudes in the 8 bands, 
$g$, $r$, $i$, $J$, $H$, $K_{\rm s}$, $W1$ and $W2$ 
are generated from the reference SED library (c.f. Fig.\,\ref{fig2})
using a random $r$-band reference magnitude $r_0$, 
ranging from 12 to 19\,mag.
Photometric errors of individual magnitudes of the eight bands are  
generated independently. Our algorithm is then applied to the resultant 
``observed'' colours to derived the best-fit values of 
$(g-i)_0$ and $A_r$. For each fiducial star, we generate the magnitudes randomly and 
calculate the best-fit parameters $(g-i)_0$ and $A_r$ of minimum $\chi^2$ 
 for 10,000 times.
Fig.\,\ref{mcfit} shows distributions of the best-fit parameters in the $(g - i)_0$ and 
$A_r$ plane, and histograms of best-fit values of $A_r$.
The top, middle and bottom panels 
correspond to the three simulated cases of intrinsic colours, 
i.e. $(g - i)_0 = 0.5$ (a blue star), 1.95 (a star near 
the ``knee'' of the stellar locus), and 2.50 (a red star), respectively.
In each row, the three panels refer to the three cases of 
photometric errors: 0.02, 0.05 and 0.08\,mag for (a), (b)
and (c), respectively. As noticed by \citet{Berry2011} earlier, the covariance of 
the best-fit values of $(g-i)_0$ and $A_r$ is larger 
for a blue star than a red one. For stars 
with high photometric accuracy ($\sigma _i$ =0.02), 
the algorithm recovers the values of $(g - i)_0$ and 
$A_r$ perfectly, a blue, ``knee'' or red star, regardless.
For a blue or red star, the  $\chi^2$ distribution  always possesses only  one peak, 
even for $\sigma _i$ = 0.08\,mag. The best-fit algorithm is quite robust in
these two cases where increasing the photometric errors simply leads to larger uncertainties
of the results. However, for a ``knee'' star, the best fit $A_r$ distribution becomes 
quite complicated, showing multiple peaks
for photometric errors larger than 0.05\,mag. 
For such stars, results yielded by the method should be treated with caution.

\begin{figure}
 \centering
   \includegraphics[width=0.48\textwidth]{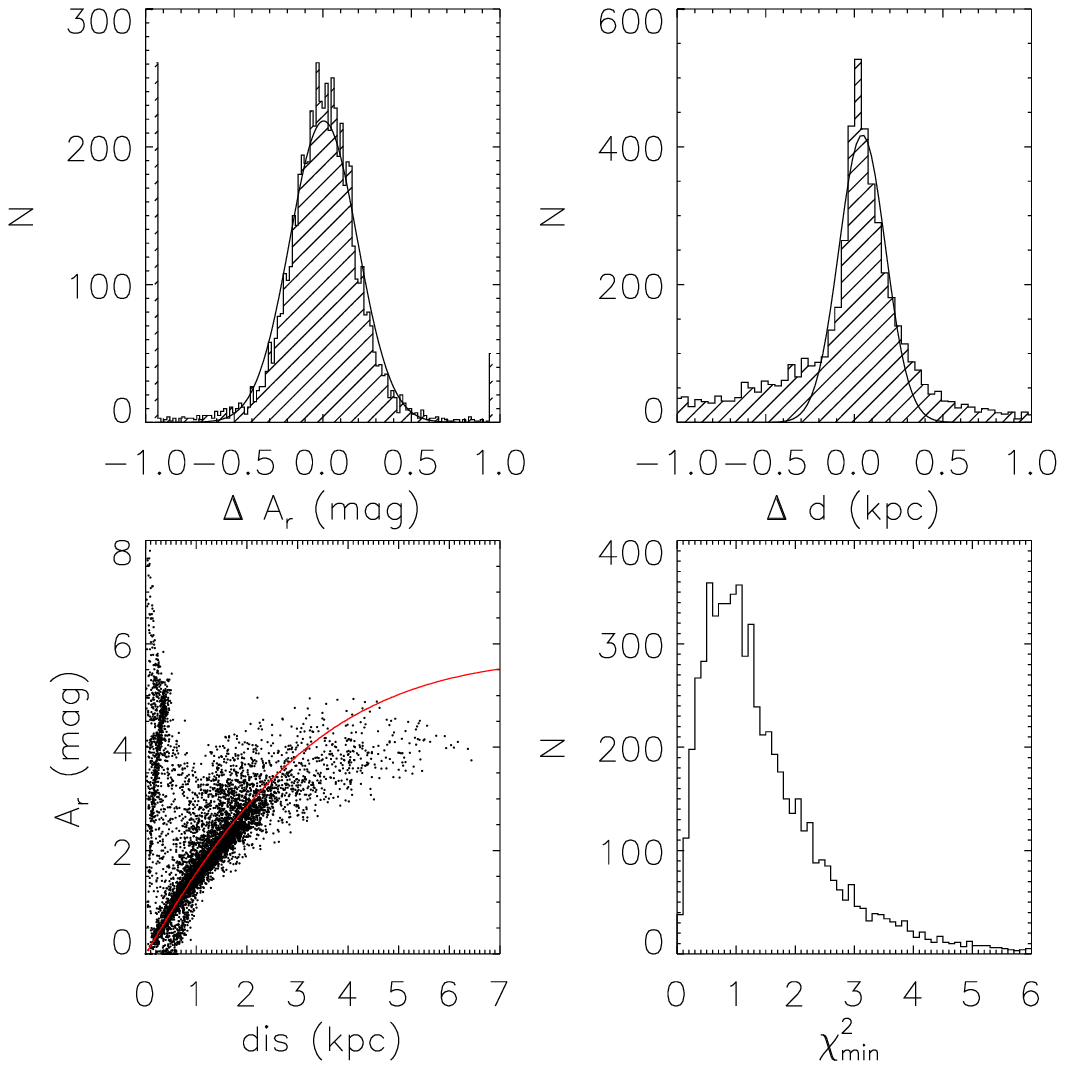}
\caption{Test results for a one squared degree field simulated with Besan\c{c}on model.
{\em Upper left:}\, Histogram distribution of differences of $A_r$ extinction values 
deduced from our fitting algorithm and those assumed in the model. {\em Upper right:}\, 
Same as upper left but for stellar distances. {\em Lower left:}\, $A_r$ plotted against 
distance. Values recovered by our fitting algorithm are represented by dots. The 
red solid line is the artificial extinction assumed in the Galactic model. 
{\em Lower right:}\, Histogram of minimum 
$\chi^2$ values of fits.}
\label{besres}
 
\end{figure}

Next we test our algorithm using mock catalogs generated with the Besan\c{c}on Galactic
stellar population synthesis model
\citep{Robin2003,Robin2012}.
The Besan\c{c}on Galaxy model is 
based on a scenario of the formation and evolution of the 
Milky Way. It simulates the stellar contents of the Galaxy with 
four distinct stellar populations: a thin disc, a thick disc, 
a bulge and a spheroid. 
It also takes into account the 
presence of a dark halo and a diffuse component of 
the interstellar medium. To test our method, we create artificial data
by adding manually extinction values to intrinsic magnitudes generated by the 
model. Fig.\,\ref{besres} shows an example of our test, for a one squared degree field 
centered at ($l,~b)~=~(180.00,~-10.00$). Since  
the current Besan\c{c}on model does not include $W1$ and $W2$ bands, we 
use the IRAC [3.6] and [4.5] bands as surrogates since they are very close. 
We assume an extinction gradient of 2.0\,mag/kpc 
in $V$ band. The completeness limit is set to 
19\,mag in $r$ band and photometric errors to 0.05\,mag for all bands. 
Note that the Besan\c{c}on model computes the photometry using 
stellar atmosphere models (Basel 3.1) from \citet{Lejeune1997} and 
\citet{Lejeune1998}. The Basel 3.1 stellar library is different from 
our empirical reference stellar library (Fig.\,\ref{fig2}). 
So we replace our reference stellar locus by the Basel 3.1 
stellar library for consistency. We also assume the empirical extinction
law of \citet{Mathis1990}, to be consistent with the Besan\c{c}on model.   

The test results are presented in 
Fig.\,\ref{besres}.
The lower left panel of Fig.\,\ref{besres} plots the
best-fit distances versus values of  
extinction recovered by our algorithm. The overplotted red line delineates the 
increase of extinction as a function of distance assumed in
the Galactic model and the black dots are the best-fit values of extinction 
and distance for individual stars of the mock catalogue. For most of the stars,
the best-fit values trace closely the model ones.
Those well behaving stars are in fact dwarfs.
There are however a bunch of stars in the 
upper left parts of the
diagram, i.e. stars at small distances yet having very 
high values of extinction. Those are giant stars. For giants, our algorithm has grossly underestimated 
their distance. Giants have intrinsic colours almost identical to dwarfs. 
However, they are intrinsically much brighter and therefore 
for a magnitude limited sample, most giants are 
located at much further away than dwarfs.  
The distributions of
differences of our best-fit and model values are well represented by a Gaussian, 
for the cases of both distance and extinction 
 (top panels of Fig.\,\ref{besres}). For extinction and distance, the 
average value and dispersion of the differences are 
respectively $-0.006\pm 0.16$\,mag and
$-0.05\pm 0.13$\,kpc. The lower right panel of Fig.\,\ref{besres}
shows the resulting minimum $\chi ^2$ distribution of the algorithm.
As expected, the $\chi ^2$ distribution peaks at unity. The distributions of distance
differences and $\chi ^2$ values are skewed because of the underestimated 
distances for giant stars.
The test shows that  our method can successfully
recover the extinction values and distances for dwarfs but underestimates the  
distances of giants.
The simulations also show that the contamination  
of giant stars in the XSTPS-GAC sample 
is quite small, probably less than 10\% \citep{Liu2013}.

The above test of fiducial stars yields residuals of extinction
 of 0.002$\pm$0.1, 0.01$\pm$0.2 and 0.02$\pm$0.3\,mag for photometric errors of 
0.02, 0.05 and 0.08\,mag, respectively, whereas 
the test with Besan\c{c}on model 
yields residuals of $-0.006\pm0.16$\,mag and $0.045\pm0.13$\,kpc
 for extinction and distance, respectively.
Both tests thus show that  except for giants our method and 
implementation of the $\chi^2$ minimization algorithm is capable of
yielding accurate parameters for most stars.
Values of reddening together with distances for individual dwarfs 
can be accurately recovered. 

\section{Result}

\subsection{$\chi ^2$ distribution }

\begin{figure}
\centering
\includegraphics[width=0.48\textwidth]{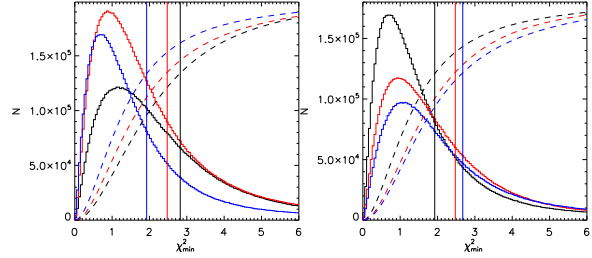}
\caption{Distributions (solid curves) and accumulative 
distributions (dashed curves) for minimum $\chi^2$ values of 
best-fit parameters for sample stars in different ranges of
 Galactic latitude (left panel; {\em Black:}\, $|b| < 5$\,deg; 
{\em Red:}\, $5 \leq |b| \leq 15$\,deg; {
\em Blue:}\, $|b| > 15$\,deg), and 
for stars with different ranges of mean photometric errors 
(right panel; {\em Black:}\, $\bar{\sigma} < 0.05$\,mag; 
{\em Red:}\, $0.05 \leq \bar{\sigma} \leq 0.08$\,mag; {\em 
Blue:}\, $\bar{\sigma} > 0.08$\,mag). The three vertical lines 
refer to the 68.3\% cut of the corresponding cumulative distributions.}
\label{fig3}

\end{figure}

Our algorithm is applied to the subset of the XSTPS-GAC, 2MASS and WISE combined 
 multi-band photometric catalog (c.f. Section\,2.4). 
We present our results by first discussing the 
distribution of minimum $\chi^2$ values of 
best-fit parameters, $\chi _{\rm min} ^2$.
Fig.\,\ref{fig3} shows the distributions of 
stars located at different ranges of Galactic 
latitude, as well as for stars with different ranges of photometric errors.
All distributions peak around unity.
As Galactic latitude decreases, stars begin to suffer 
from higher extinction,
have higher photometric errors, and are contaminated by
more giants, all these make the $\chi^2_{\rm min}$  distribution shifting toward 
larger values. 
For the three regions of different Galactic latitude range plotted in the 
left panel of Fig.\,7, the 
68.3\% cut of 
the accumulative distributions correspond to $\chi^2_{\rm min}$ values of 2.85, 
2.45 and 1.95 for regions of $|b| > 15$\,deg, $5 \leq |b| \leq 15$\,deg and 
$|b| < 5$\,deg, respectively.

The right panel of Fig.\,\ref{fig3} shows $\chi^2_{\rm min}$
distributions for groups of  
stars of different ranges of photometric errors. We divide 
the sample into three groups by the average values of 
photometric errors of the eight bands:
$0.02~ \leq~ \bar{\sigma}~ <~0.05$\,mag,
 $0.05~ \leq~ \bar{\sigma}~ \leq~0.08$\,mag and $\bar{\sigma}~ >~0.08$\,mag,
representing in sequence stars of most to least reliable photometry.
Note that we have imposed an $r$-band limiting magnitude cut $r~<~$18.5\,mag 
when defining the sample (c.f. Section\,2.4), implying that while 
most stars have $r$-band errors less than 0.05\,mag, there are some stars
 in the sample  whose
$r$-band or other bands errors are larger than 0.08\,mag. Also note that we have set
the minimum photometric errors to 0.02 for all eight bands
to account for possible systematic (calibration) uncertainties.
Fig.\,\ref{fig3} clearly shows that as photometric errors increase, 
the $\chi^2_{\rm min}$ distribution shifts to larger values.
The $\chi^2_{\rm min}$ values corresponding to the 68.3\% cut of
the accumulative distributions are 1.95, 2.45 and 2.70, respectively, 
for the three groups of stars of increasing mean photometric errors.

Stars with $\chi^2_{\rm min}<2.0$ are included in our 
catalog of final results. The catalog thus contains
more than 68.3\% stars located at 
high Galactic latitudes $|b| > 15$\,deg
and the same fraction of stars of mean 
photometric errors $\bar{\sigma}$ less than 0.05\,mag.
The catalog consists of more than 14.6 million stars.
In the catalog, stars with mean photometric errors 
$0.02~ \leq~ \bar{\sigma}~ <~0.05$\,mag are flagged by ``A'' for most reliable results,
 stars with $0.05~ \leq~ \bar{\sigma}~\leq~0.08$\,mag by ``B'', 
and those with 
$ \bar{\sigma}~ >~0.08$\,mag by ``C'' for least reliable results. 
For convenience of  analysis below,
we define Sample\,A that contains all stars with flag ``A''
in the catalog of final results and Sample\,C that includes
all stars in the whole catalog of final results. Note that 
Sample\,C does not only include the stars with flag ``C''
but also with ``A'' and ``B'' as well.

\subsection{The three dimensional distribution of dust}

\begin{figure*}
\centering
   \includegraphics[width=\textwidth]{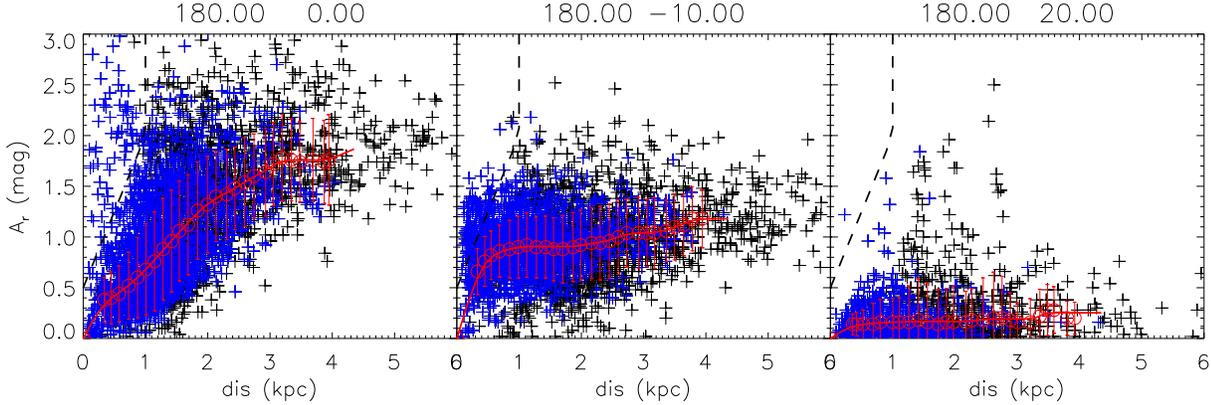}
\caption{Extinction plotted against distance for stars in three example fields
 of $1^\circ \times 1^\circ$.
The central Galactic longitude and
latitude are marked on top of each panel. Black and blue 
symbols denote stars from Sample\,C
and A, respectively. The red circles and errorbars
are running medians and standard deviations of 
all data points of Sample\,C, excluding those above the dashed lines.
The latter are most likely giants whose distances may have been grossly underestimated. 
The red lines are smooth delineate the median values after applying a boxcar smoothing. 
See text for more detail.}
\label{los3d}

\end{figure*}

The catalog of final results, containing best-fit $A_r$ and 
distance parameters for over 11\,million stars, is publicly 
available online, along with the reference 
stellar locus library\footnote{http://162.105.156.249/site/Photometric-Extinctions-and-Distances.}. 
As illustration, in Fig.\,\ref{los3d} the values of extinction are plotted against 
distances for stars in the final catalog for
 three sample $1^\circ \times 1^\circ$ fields centered at 
$l=180^{\circ}$ and $b$=0$^{\circ}$, $-10^{\circ}$
and 20$^{\circ}$).
Not surprisingly, most stars of Sample\,A are located  at closer
distances. This is because Sample\,A stars 
have higher photometric accuracy, and are on average brighter and 
thus at smaller distances. 
The dashed line in each panel marks the 
``boundary'' between dwarfs and giants \citep{Berry2011}. 
Stars above the dashed lines are likely giants with grossly 
underestimated distances (c.f. Section\,3.5 and Fig.\,6).
There are approximately 2 million stars above the dash lines in total, 
 less than 20\% of the whole final sample.
The same criteria for the selection of giant candidates with
Berry et al. (2012, Section.\,3.1.5) is adopted for the entire catalog
of the final results and 75,541 stars are selected out as 
giant candidates. These candidates are
cross-matched with the LSS-GAC data (Yuan et al. 2014, to be submitted)
and we obtain the basic parameters, $T_{eff},~{\rm [Fe/H]~,log}g$, for
 4,929 stars of them. About 80\% of the 4,929
stars have ${\rm log}\,g$ smaller than 4.0 and have a 
peak at ${\rm log}\,g~=~2.6$, implying that the giant candidates selection is
quite reliable. We mark these candidates with an additional flag
as ``G'' in the final catalog.    

In principle, one should see that extinction increases with distance.
Such a trend is not so obvious for high latitude fields, 
given the low extinction of those fields and the fact that for those fields most stars
are likely behind the dust layer. On the other hand, the trend is very significant for low latitude fields.
As pointed out earlier, in the plane, most of the stars  are embedded within
the dust layers  rather than behind it. Thus if one uses the total extinction 
from a 2D extinction map (such as that of SFD)  to deredden a disk star,
it is quite likely that the extinction has been overestimated.
The analyses of \citet{Arenou1992} and \citet{Chen1999} assumed that
extinction as a function of distance can be approximated by a 4th order 
 polynomial, $E(B-V)=a_1r+
a_2r^2+a_3r^3+a_4r^4$. 
However studies including the current one 
\citep{Marshall2006,Chen2013,
Schultheis2013}  have shown that the relation between extinction and distance is quite complicated 
and varies from field to field. The results presented in Fig.\,\ref{los3d}
illustrate well the power of a 3D extinction map. Firstly, the extinction in the field 
$(l,b)=(180^\circ,0^\circ)$ increases with distance up to a distance of $\sim$4\,kpc, while for 
the field $(l,b)=(180^\circ, 20^\circ)$ the extinction stops increasing at a distance of only 0.3\,kpc.
The results thus indicate a scale length and height of the dust layer of less than 4\,kpc 
and 0.3\,kpc, respectively. Secondly, measurements of the 3D  
dust distribution can be used to detect and trace dark clouds
\citep{Marshall2009, 
Schlafly2014}. For example, in the field $(l,b)=(180^\circ,-10^\circ)$, 
one sees a dramatic 
increase of the extinction at a distance of about 0.2\,kpc, which is 
in fact the location of the well known 
Tauras dark cloud \citep{Straizys1992}. Finally, a 3D extinction 
map of large sky area can trace the large structure of the dust distribution. 
\citet{Schultheis2013} find that the extinction first increases significantly at about 5--7\,kpc
and then decreases after 8\,kpc toward the Galactic center. They suggest that there could exist
a dust lane in front of the Galactic bar.

\begin{figure*}
\centering
   \includegraphics[height=0.8\textheight]{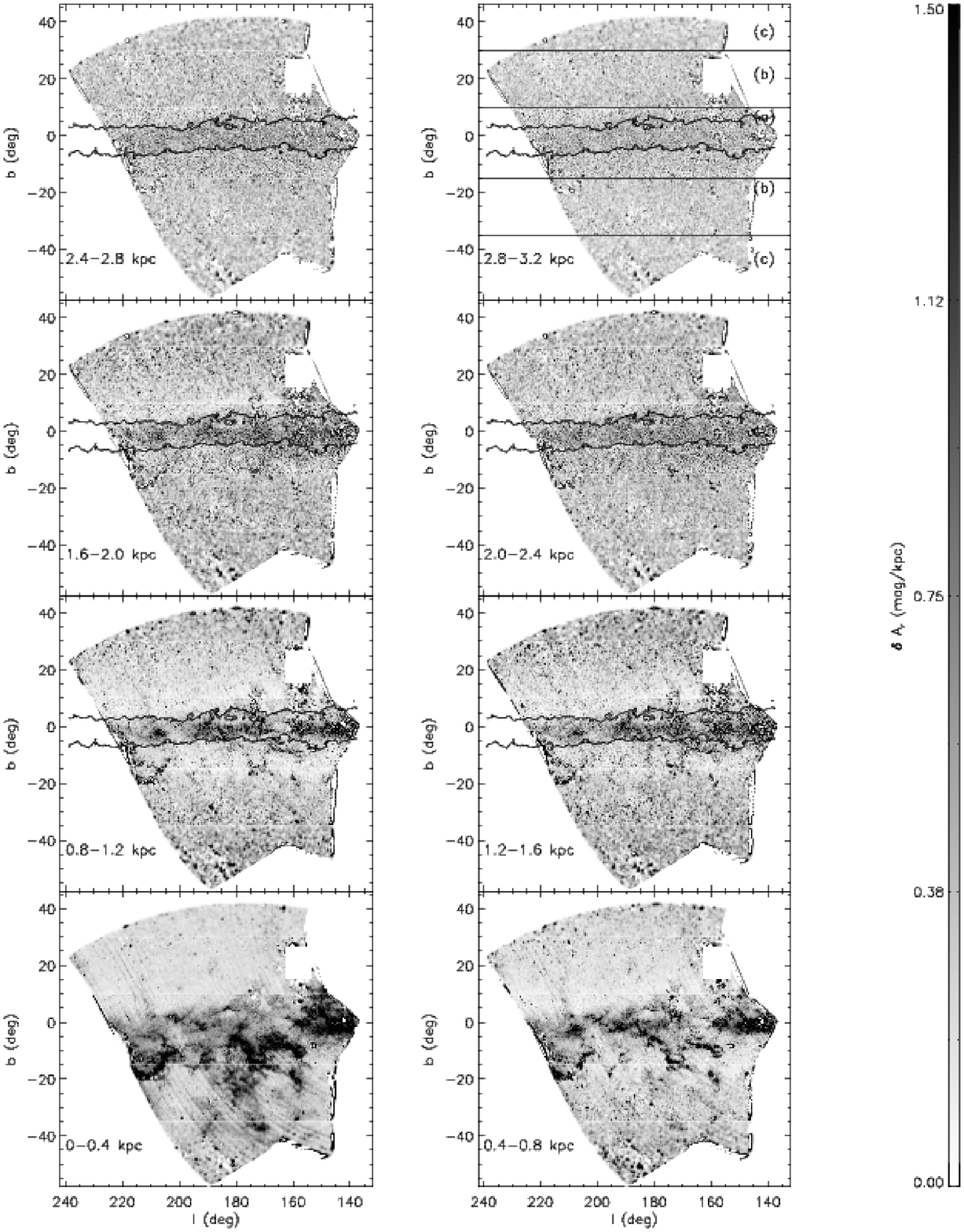}
\caption{Extinction of individual distance bins of length 400\,pc. 
The distance interval the extinction map refers to is marked in each panel.
Different angular resolutions are adopted for regions 
 of different Galactic latitudes, as marked 
in the top right panel, and are 
3, 6 and 9\,arcmin for regions (a), (b) and (c), 
respectively (see text for details). The contours overplotted in panels of large distance bins
($d>0.8\,kpc$) are integrated emission of H\,{\sc i} gas from the LAB all-sky
survey \citep{Kalberla2005}.}
\label{map3d}

\end{figure*}

We employ a sliding window of width 450\,pc with a step of
150\,pc to obtain median value of extinction for each distance bin, 
excluding stars above the dashed lines, which are likely giants 
rather than dwarfs. The red lines in 
Fig.\,\ref{los3d} plot the median extinction 
as a function of distance for the three fields 
after applying a boxcar smoothing of a width 450\,pc. 
We group all stars in the catalog of final results into subfields
of varying angular dimensions and
calculate the median value of extinction in each sliding window, 
again excluding possible giants. 
The results are then used to
derive extinction as a function of distance for each subfield.
The adopted dimension of subfields (angular resolution) of the 
resultant map depends on latitude, as shown in the top right 
panel of Fig.\,\ref{map3d}. For latitudes ranging $-15~<~b~<$ $+10^\circ$
[denoted as region (a) in the panel], the stellar  
density is high enough to allow a high angular resolution of 3\,arcmin. 
As the stellar density drops towards  high Galactic latitudes,
we reduce the resolution to 6\,arcmin for $-35$ $\leq~ b ~\leq$ $-15^\circ$ and 
$10$ $\leq~ b ~\leq$ $30^\circ$ [regions (b)], 
and to 9\,arcmin for $b$ $<$ $-35^\circ$ and $b$ $>$ $30^\circ$ [regions (c)]. 
For the high angular resolution regions, we require that there are at least 10
stars in each bin.
Fig.\,\ref{map3d} shows resultant 3D extinction maps for
the entire GAC area  surveyed by the XSTPS-GAC. 
The individual panels in the diagram present 
extinction maps by local dust grains in individual 
distance bins of length 400\,pc 
(0 -- 0.4, 0.4 -- 0.8, 0.8 -- 1.2, 1.2 -- 1.6, 1.6 -- 2.0, 2.0 -- 2.4, 2.4 -- 2.8
and 2.8 -- 3.2\,kpc).

Fig.\,\ref{map3d} represents the very first set of high angular resolution 3D extinction maps;
covering about 6,000\,$\rm deg^2$ sky area of the GAC and spanning a distance range
from 0 to 4\,kpc. The data are available online from the aforementioned website. They
are also available in electronic form at
the  CDS\footnote{The data is only available in electronic form at the CDS via anonymous 
ftp to cdsarc.u-strasbg.fr (130.79.128.5) or via 
http://cdsweb.u-strasbg.fr/cgi-bin/qcat?J/A+A/.}. Table\,2 provides an example of the format. 
Each row of Table\,2 contains the information for one line of sight: Galactic coordinates along 
with the measured quantities for each distance bin $A_r$, and the respective uncertainties.
The width of the distance bin is 150\,pc.

\begin{table*}
 \centering
  \caption{Resulting extinction as a function of Galactic longitude, latitude, and distance.}
  \begin{tabular}{cccc}
  \hline
  \hline
$l$ (deg) & $b$ (deg) & $A_{r,0.0-4.35kpc}$ (mag) & $\sigma A_{r,0.0-4.35kpc}$ (mag)  \\
 \hline
\end{tabular}\\
Notes. For each position we provide the $r$\,band extinction, and the corresponding sigma for 
each distance bin starting from 0.0 to 4.35 kpc, with a step of 150\,pc.
  \label{ta2}
\end{table*}

Many interesting features are clearly visible in the dust distributions presented in 
Fig.\,\ref{map3d}, including the  
obvious disk warps.
The features  seen in Fig.\,\ref{map3d} are  consistent with previous 
studies \citep{Schlegel1998, Dobashi2005, Froebrich2007}
as well as with results from studies of the gas component 
[e.g. H\,{\sc i} gas from \citet{Bajaja2005}, CO gas from \citet{Dame2001}]. 
The large scale features in this sky region, such 
as $\rm \lambda$-Ori around
 Galactic longitude and latitude
(189$^{\circ}$,$-$16$^{\circ}$), Orion A around 
(205$^{\circ}$,$-$14$^{\circ}$), Orion B around 
(213$^{\circ}$,$-$20$^{\circ}$), Auriga 1 around 
(166$^{\circ}$,$-$8$^{\circ}$), Auriga 2 around 
(180$^{\circ}$,$-$7$^{\circ}$), Taurus around 
(170$^{\circ}$,$-$15$^{\circ}$), Camelopardalis 1 around 
(150$^{\circ}$,0$^{\circ}$), Camelopardalis 2 around 
(158$^{\circ}$,0$^{\circ}$), Monoceros around 
(220$^{\circ}$,$-$5$^{\circ}$), Perseus around 
(160$^{\circ}$,$-$20$^{\circ}$), Taurusextended 1 around 
(180$^{\circ}$,$-$25$^{\circ}$), Taurusextended 2 around 
(172$^{\circ}$,$-$35$^{\circ}$), Aries around 
(160$^{\circ}$,$-$30$^{\circ}$), and Eridanus around 
(190$^{\circ}$,$-$35$^{\circ}$), are clearly visible in the local volume, within 800\,pc. 
Beyond this distance, we see mainly features from the Galactic 
thin disk.
The contours overplotted in maps beyond 0.8\,kpc show the  distributions
of integrated intensity of H\,{\sc i} gas from Leiden-Argentina-Bonn (LAB) 
21\,cm H\,{\sc i} emission survey \citep{Kalberla2005}. The features revealed by the dust 
are very similar to those of the gas.
 At distances beyond 3\,kpc, 
our results become less reliable, due to less numbers of stars
with high quality photometry available for extinction determinations.
There is a notable empty patch around $l \sim$160$\degr$ and $b \sim$20$\degr$,
due to abnormally large photometric errors of XSTPS-GAC.

\begin{figure}
\centering
   \includegraphics[width=0.45\textwidth]{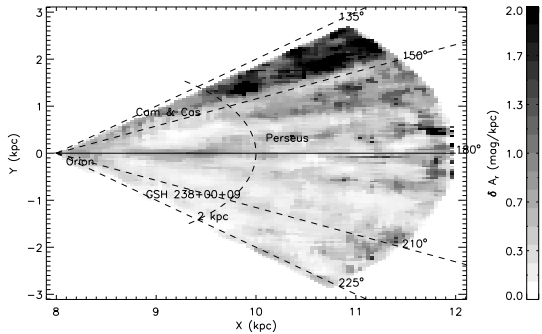}
\caption{Distribution of 
differential extinction (in units of mag\,kpc$^{-1}$) in the 
plane of  Galactic disk derived from stars of vertical distances from the 
plane less than 150\,pc. The solid horizontal line denotes Galactic 
longitude $l = 180^\circ$, and the four dashed lines denote, from bottom 
to top, longitudes $l = 225$, 210, 150 and 135$^\circ$, respectively. An arc of 2\,kpc distance
is also plotted marking the approximate position of the Perseus arm. }
\label{galz}

\end{figure}

Fig.\,\ref{galz} presents the extinction distribution in 
the Galactic disk plane from stars of vertical distances 
from the disk plane $|Z|$  $<$ 150\,pc. 
It gives the extinction by local dust  (in units of mag/kpc). The Sun is 
located at
 $X=8$\,kpc and $Y=0$\,kpc. 
A number of features are visible  in the figure, including the local Orion arm that 
extends to about 800\,pc at longitude of  about\,210$^{\circ}$. The Perseus arm can been 
seen at distances of  about\,2--3\,kpc. The well known  Camelopardalis and  Cassiopeia 
dust features are visible 
at longitudes between  130$^{\circ}$ and 150$^{\circ}$ 
beyond 800\,pc. The position of the features are consistent with 
the location of young stars found  in the Camelopardalis dust and molecular clouds 
\citep{Straizys2008}. 
%The figure also shows an  inverted 
%differential opacity distribution in the local volume, 
%similar to that found earlier by  \citet{Lallement2013} in local area.
The super-bubble GSH 238+00+09 detected in the radio by \citet{Heiles1998}
is visible at a longitude of  about 220$^{\circ}$. 

\begin{figure}
\centering
   \includegraphics[width=0.48\textwidth]{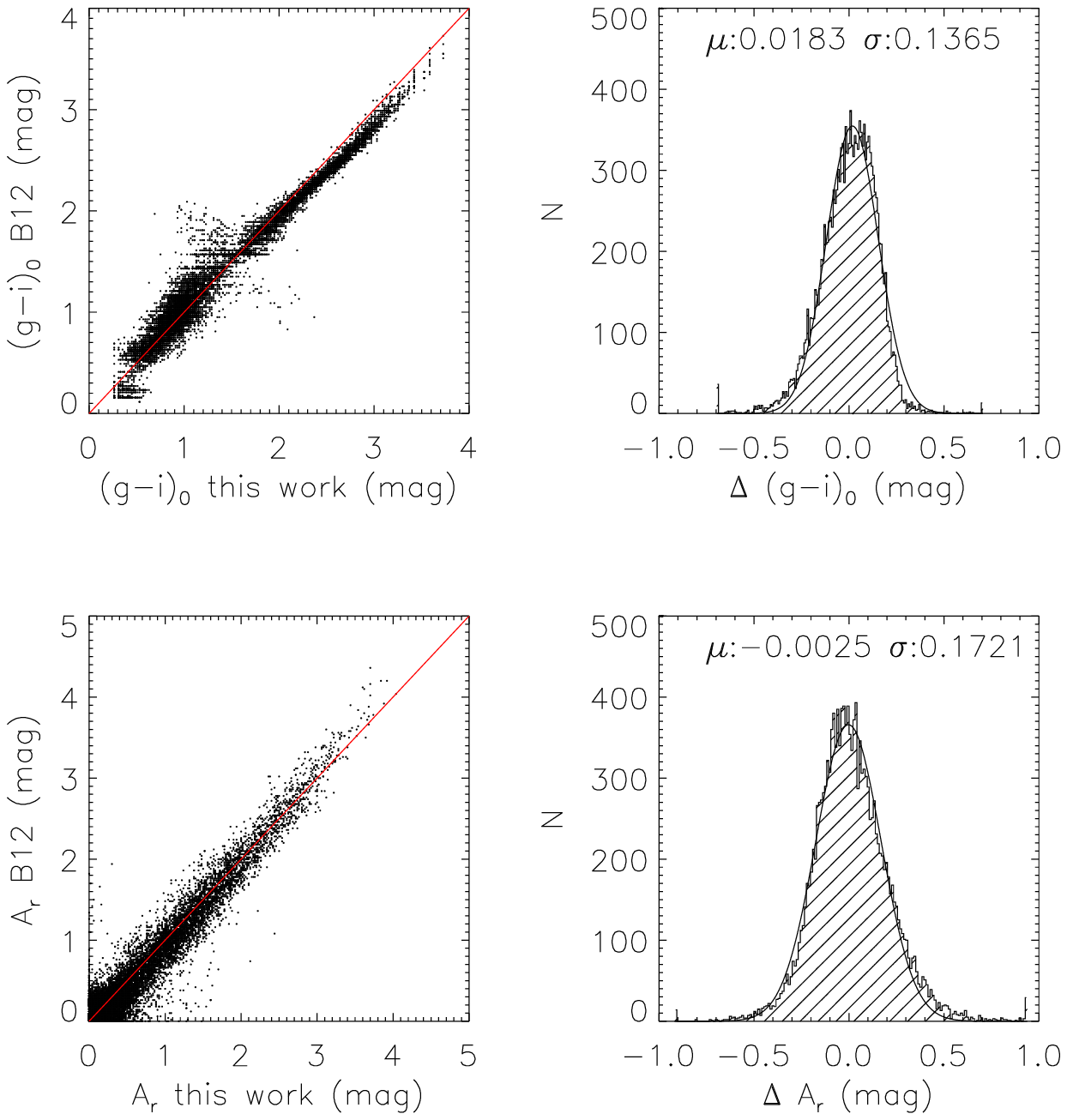}
\caption{Comparison of best-fit values of 
$(g-i)_0$ (upper left panel) and $A_r$ (lower left panel) from
\citet{Berry2011} and the current work for common objects.
 Histograms of differences of the two sets of determinations are plotted in the 
right panels, for $(r-i)_0$ (top right) and $A_r$ (bottom right), respectively.}
\label{compbry}

\end{figure}

\citet{Berry2011} carry out a similar analysis for unresolved sources in the
SDSS DR7 release 
based on  SDSS  (and 2MASS  when available) photometry. Their analysis include data 
collected as parts of the SEGUE, which images ten 2.5\,deg wide 
stripes that cross the Galactic plane at selected longitudes. Some of those stripes 
overlap with the footprint of XSTPS-GAC. For comparison, we have cross-identified 
common targets analyzed by both \citet{Berry2011} and the current work. 
We use their results deduced with both the  SDSS and 2MASS photometry 
available and with $R_V$ fixed  3.0. The cross-identification is carried out 
 using a matching radius of 1.5$''$. 
Stars in \citet{Berry2011} with best-fit $\chi ^2 ~<~2$, $r~ <~19$\,mag and $K_{\rm s} < 14$\,mag 
are selected for comparison with our results. The 
results are shown in Fig.\,\ref{compbry}. In spite of  different set of optical photometry  and 
reference stellar locus  library adopted in the two studies, 
we see  rather good agreement between their results and ours. 
Generally, the differences are well fitted by a Gaussian function 
(right panels of Fig.\,\ref{compbry}).
With a mean and standard deviation of $-0.003\pm 0.170$\,mag, 
there are no systematical differences between the two sets of determinations of extinction.  For
$(g - i)_0$,  the scatter is a bit large for stars located on the blue portion 
of the stellar locus (c.f. Fig.\,\ref{fig2}), and to a less extent, for stars located near 
the ``knee'' of the stellar locus. 
An opposite twist is found for the extinction, although not so obvious.
The twist is caused mainly by the different sets of stellar 
locus used and possibly in parts due to 
the degeneracy between the intrinsic colours and reddening.
In fact, if we recalculate the best fit values of extinction 
$A_r$ and intrinsic color $(g-i)_0$ using the SDSS stellar locus 
(Davenport et al. 2014) instead of ours, then the twist  seen in the upper 
left panel of Fig.\,11 becomes largely absent.

\subsection{Two dimensional extinction maps}

We construct  the 2D extinction map using two methods. The first is using the 
best-fit $A_r$ values from  our catalog of final results. The other is to use the
Rayleigh–-Jeans Color Excess (“RJCE”) method
proposed by  \citet{Majewski2011} based on  the 
2MASS and WISE IR photometry only. We first describe the results from the 
 RJCE method. 

\begin{figure}
\centering
   \includegraphics[width=0.48\textwidth]{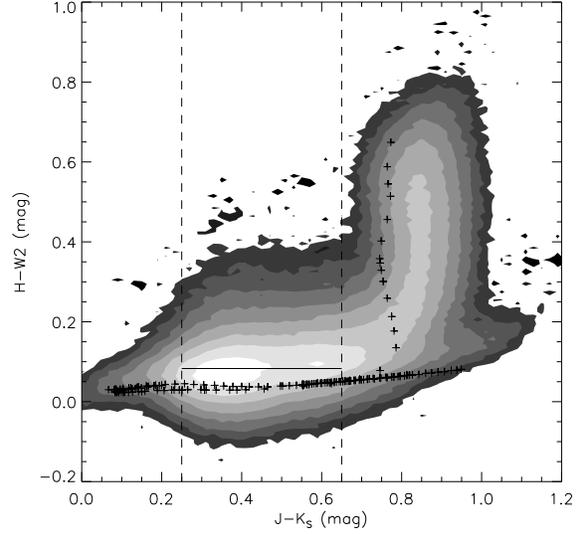}
\caption{$ (J-K_{\rm s})$ - $(H - W2)$ colour-colour 
diagram for the high-quality reference sample consisting of 132,316 stars 
of very low extinction. Pluses delineate an isochrone of age 10\,Gyr 
and Solar metallicity from \citet{Girardi2002}. The $(J - K_{\rm s})$ colours range [0.25, 0.65]  
is marked by two vertical dashed lines. 
The black horizontal line indicates a mean of 0.083\,mag of $(H-W2)$ colour for stars
of $(J - K_{\rm s})$ colours in the range.}
\label{wscl}

\end{figure}

\begin{figure*}
\centering
   \includegraphics[width=\textwidth]{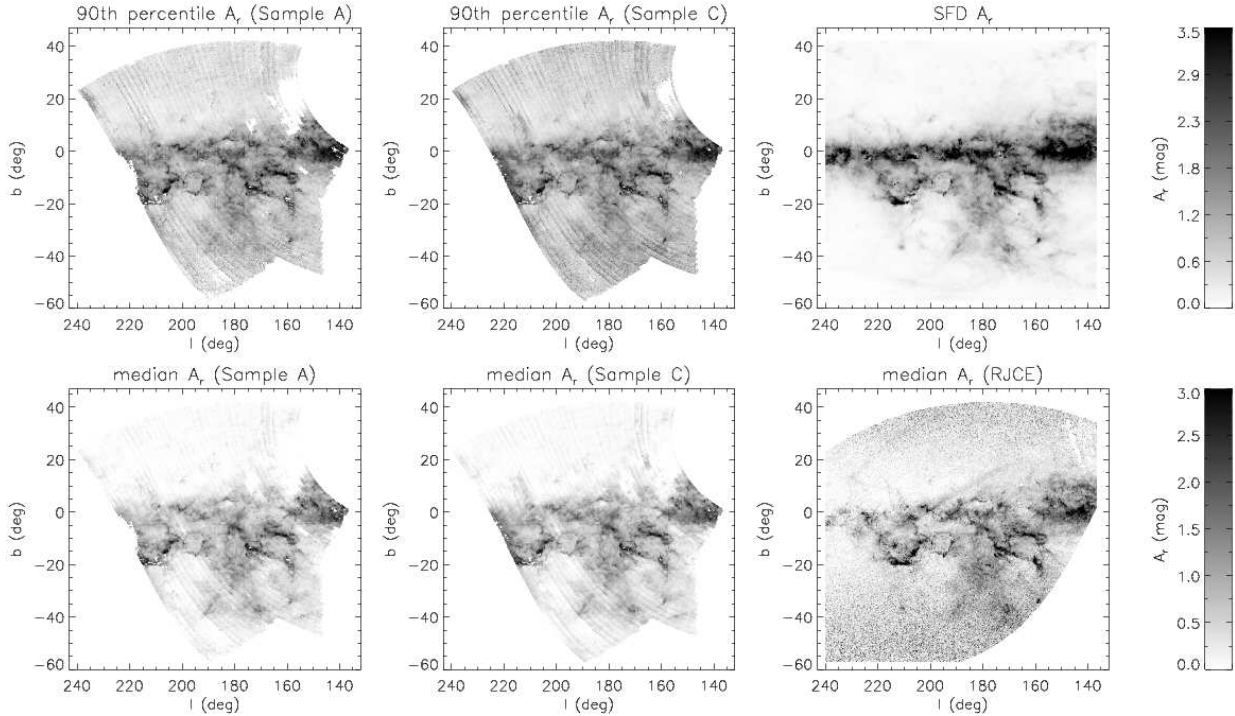}
\caption{Two dimensional extinction maps. The top  three show maps 
generated from the  90th percentile of 
extinction  of Sample\,A (left) and C (middle), and that from  SFD (right). The bottom three
show maps generated from the  median values of extinction of Sample\,A (left) and C (middle), 
and that generated from  
the median values of  extinction derived using the  RJCE method.} 
\label{fig11}

\end{figure*}

For the RJCE method, we use colour $(H-W2)$ to estimate extinction. 
In Fig.\,\ref{wscl}, 
we show the $(H-W2)$ v.s. $(J - K_{\rm s})$ colour-colour diagram for the 
high quality reference sample consisting of 132,316 stars of very 
low extinction (see Sect.~2). Also overplotted is an isochrone of stars of 
age 10\,Gyr and solar metallicity $Z = 0.019$ from  \citet{Girardi2002}.  
Stars  in the colour  range $0.25 ~ \leq ~ (J-K_{\rm s})~ \leq ~ 0.65$\,mag, 
mostly main sequence dwarfs plus a few red clump giants  have nearly 
constant $(H - W2)$ colour, 
with an average and standard deviation value
of 0.083 and 0.039\,mag, respectively.   
Accordingly,  we set the zero point of $(H - W2)$  colour to
$(H-W2)_0=0.08$\,mag. 
Then from the extinction law of \citet{Yuan2013}, we have,
\begin{equation}
A_r=7.484 \times (H-W2-0.08)
\end{equation}
We select  stars in the 2MASS and WISE catalogs that are detected in 
all $J$, $H$, $K_{\rm s}$, $W1$ and $W2$ bands and photometric errors
$<$ 0.08\,mag, in the sky area, 
 3h $<$ R.A. $<$ 9h and $-$90  $<$ Dec. $<$ 90\,deg, 
and calculate their extinction  values using the above equation.  The 
stars are then dereddened using the derived values of extinction 
and only  stars  with dereddened $(J - K_{\rm s})$ colours,  
$0.25 ~ \leq ~ (J-K_{\rm s})_0~ \leq ~ 0.65$  are retained. 
We then bin the stars  into subfields, each of angular size 6$'~\times~$6$'$.
A bootstrap algorithm is used to calculate the median and 
standard deviation (s.d.)  values of  extinction 
of all  stars in a given subfield. The resultant 2D extinction map is shown in   
the bottom right panel of Fig.\,\ref{fig11}.

In Fig.\,\ref{fig11} we also show the 2D extinction maps generated  
from our final catalog that contains
best-fit values of extinction  $A_r$. Again we bin stars in both  
Sample\,A and C separately into 
subfields of size $6'~ \times ~ 6'$. A bootstrap algorithm 
is applied to calculate
the median and s.d. values of extinction for stars
in each subfield. Meanwhile the value of the 90th percentile of extinction  and the
 associated s.d. of  each 
subfield are also calculated with the same algorithm.  
The results are also plotted in Fig~\ref{fig11}. 
For comparison, Fig.\,\ref{fig11} also show
the SFD extinction map in the upper right panel of the diagram. 

Compared to Sample\,C, due to the fewer number  of stars available from 
Sample\,A, there are more ``empty'' subfields  in the 
Sample\,A extinction maps than those of Sample\,C, 
in particular near the edges of footprint of the XSTPS-GAC. 
Also there is a notable empty patch  around $l$
$\sim$ 160$^{\circ}$ and $b$ $\sim$ 20$^{\circ}$, due to abnormally 
large photometric errors of XSTPS-GAC fields in this area. 
Similar ``holes'' at the same location are also apparent in Fig.\,\ref{map3d}.  
There are also some ``stripe-like'' structures in the best-fit extinction maps. 
This is caused by the nightly varying survey depth and photometric accuracy of 
XSTPS-GAC. Overall, the extinction map obtained with the RJCE method agrees 
well with those from the best-fit SED algorithm (c.f. bottom panels of Fig.\,\ref{fig11}). 
Similarly, there is high degree of similarity of the overall structure and 
features between the best-fit 90th percentile extinction maps and that of 
SFD (top panels of Fig.\,\ref{fig11}).  
All dust features visible in the
3D maps (Fig.\,\ref{map3d}) are clearly seen in these 2D maps.

To a good approximation, the 90th percentile (which exclude most of the extreme outliers) 
 extinction maps 
constructed from our best-fit Samples\,A and C represent most 
of the extinction along individual lines of sight and thus can be treated as 
a ``total'' extinction map for the sky area covered. Compared to the
SFD map, our ``total'' extinction maps show larger values for regions of high latitudes
($|b|~ > ~15^{\circ} $), in particular the map from Sample\,C. 
 One possible cause of this bias could be photometric 
errors.  Another potential source of errors 
is the  colour-reddening degeneracy, especially considering the small
amount of  extinction at high Galactic
latitudes, typically only a factor 2--3 larger than the photometric errors \citep{Berry2011}. 
At low Galactic latitudes where the extinction is much larger than the 
photometric errors, one sees that Sample C produces the same 
high quantity extinction map as Sample A. 
At low latitudes, the SFD map yields higher extinction
than ours. This may partly due to fact that the survey depth of XSTPS-GAC 
is not deep enough to penetrate the disk to its outer edge, in particular along 
sight lines of high extinction.

\begin{figure}
\centering
   \includegraphics[width=0.48\textwidth]{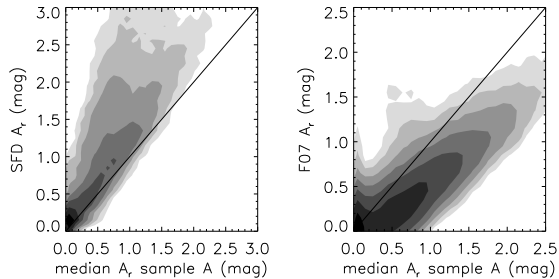}
\caption{Values of extinction given by our  Sample\,A 
median extinction map for individual sight lines are compared 
those predicted by the 2D maps of
SFD (left) and  Froebrich et al. (2007; right). 
In both panels, the straight line represents the curve of identity.} 
\label{otherm}

\end{figure}

Fig.\,\ref{otherm} compares values of extinction given by our Sample\,A  median 
extinction to those of the previous 2D extinction maps: that of SFD 
in the upper panel and of \citet{Froebrich2007}  in the lower panel. The SFD map 
gives the total extinction integrated to infinity. Therefore, for a given line of sight, 
the extinction yielded by the SFD map should in general be 
larger than our median value map.  On the other hand, 
\citet{Arce1999} argue that the SFD  map overestimated the extinction by 20--40\% 
for the Taurus region. \citet{Dobashi2005} conclude that SFD overestimated 
the extinction value by at least a factor of two, in consistent with 
what found here.

\begin{figure}
\centering
   \includegraphics[width=0.48\textwidth]{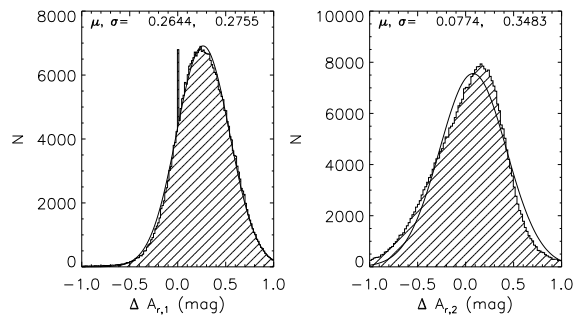}
\caption{{\em Left panel: }\,Comparisons between of $A_r$ 
values given by the median values of Sample\,A of the current work and those
of Froebrich et al. (2007). {\em Right panel: }\, Comparison of $A_r$ values from the 
integrated $A_r$ map of Sample\,A  of the current work, integrated to a 
distance of 400\,pc
and those of  Froebrich et al., after recalibrated the latter 
using the extinction coefficients of Yuan et al. (2013).}
\label{otherm2}
\end{figure}

Values obtained in the current work are systematically higher, 
by 0.2 -- 0.3\,mag than those obtained by  \citet{Froebrich2007}, 
who deduced their results using  the near-IR colour excess 
(Fig.\,\ref{otherm2}). 
Note that Froebrich et al. use the extinction law of Mathis \& Cardelli (1990) to 
convert values of $E(J-H)$ and $E(H-K_{\rm s})$ colour excess
to values of visual extinction at $V$-band, $A_V$. 
For the \citet{Mathis1990} extinction law, one has $A_V~=~10.45E(J-H)$, while for that of 
\citet{Yuan2013} one gets $A_V~=~11.92E(J-H)$. Thus applying the reddening law of 
\citet{Yuan2013} to the data of \citet{Froebrich2007}  will  reduce the systematic 
differences between their results and ours, but not quite enough. Another factor that 
may be contributing to the systematic discrepancy is that 
\citet{Froebrich2007} concentrate their  study on dust clouds (e.g. 
 Orion, Perseus, Taurus, Auriga, Monoceros, Camelopardalis, Cassiopeia), mostly 
located within 1\,kpc from us (see Fig.\,\ref{map3d}). So maybe it is not surprising that 
their reddening values  are systematically lower  than ours. 
In the right panel of Fig.\,\ref{otherm2}, we compare the extinction values between 
 $A_r$ of \citet{Froebrich2007}, after recalibrated using the extinction coefficients from Yuan et al., 
with those from our integrated map of
distance 400\,pc within which most of the dust clouds locate. 
Now the two sets of determinations are 
in good agreement, with
an average  difference of only $0.08\pm0.35$\,mag.

\section{Discussion}

\subsection{Comparison of the RJCE and best-fit methods}

\citet{Majewski2011} argue that the intrinsic $(H-W2)$ colour is almost 
independent of spectral type and luminosity class
since it measures the slope of the SED in the Rayleigh--Jeans part of the spectrum 
(see Fig.\,2 of \citealt{Majewski2011}). 
Fig.\,12 shows that the distribution of extinction free 
$(H-W2)$ colour of dwarfs and portions of the 
RC stars is well represented by  Gaussian distribution with a mean of $0.08$\,mag and
a dispersion of only $0.04$\,mag. For red giant stars [$(J-K_{\rm s})$
$\sim$ 0.8\,mag], the scatter of  $(H-W2)$ is much larger. The average value (i.e. zero
point) of $(H-W2)$ also varies. Thus in the current work, we only apply the RJCE method to
stars with  intrinsic $(J - K_{\rm s})$ colours, 0.25 $<~ (J-K_{\rm s})_0 ~<$ 0.65\,mag, 
although this RJCE method has been previously 
applied to giants and the RC stars by \citet{Nidever2012}.

\begin{figure}
\centering
   \includegraphics[width=0.48\textwidth]{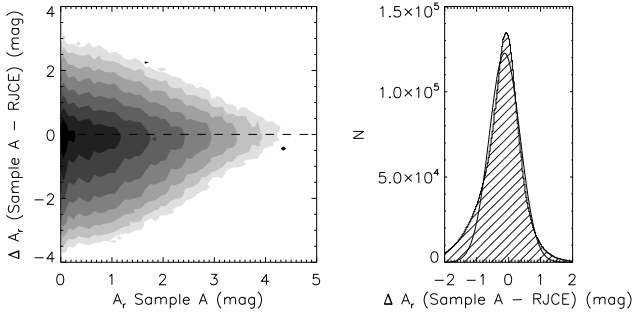}
\caption{Difference of extinction values yielded by 
best-fit SED and RJCE methods plotted against extinction (left panel). 
A histogram of the differences is given in the right panel. 
The dashed line in the left panel is 
added to guide the eye only while the black line in the right panel is a Gaussian fit
to the distribution.}
\label{comprjc}

\end{figure}

Fig.\,\ref{fig11} shows that the extinction map obtained with the RJCE
agrees with that by the best-fit SED algorithm. How do the two 
methods compare for individual stars? 
For comparison, we select stars with intrinsic colour 
0.25 $<~ (J-K_{\rm s})_0~ <$ 0.65\,mag from Sample\,A and calculate their RJCE extinction
using Eq.~(3), and compare the results for individual
stars in Fig.\,\ref{comprjc}.  
The values of difference yielded by the two methods show a roughly Gaussian 
distribution with a mean close to zero but a very large dispersion of 0.5\,mag.
 This large dispersion is found to be caused  mainly by the large 
 uncertainties for values yielded by the RJCE method. 
The RJCE makes use of 
only  one colour [i.e. $(H-W2)$], while the best-fit
SED algorithm employed in the current work uses seven colours. 
%Thus the RJCE method is much more sensitive to photometric errors than
%the best-fit SED method.     
With more colours to constrain the results, 
the best-fit SED method 
obtains more robust results
than the RJCE method does.
This is particularly true for low extinction regions, 
where typical values of extinction are comparable to the photometric uncertainties. 
The fact that we see a dramatic increase of dispersion of values of difference for 
lower values of extinction (c.f. left panel of Fig.\,\ref{comprjc}) demonstrate this.  
Note also the source confusion that WISE may be suffered from  
close to the Galactic plane may also affect the reliability of 
RJCE results significantly. Given those uncertainties,  the 90th percentile extinction 
 map deduced with the RJCE method is not shown in Fig.\,\ref{fig11}. 
 Nevertheless, the RJCE method provides a useful approach to derive 2D 
 extinction maps, especially when large numbers of stars of high 
 photometric quality are available \citet{Nidever2012}. 

\subsection{Applications of the 3D extinction maps}

Comparisons with the observed colour-magnitude diagrams 
(CMD) serves as  an important test of Galaxy stellar population 
synthesis models. 
\citet{Uttenthaler2012} and \citet{Chen2013}
 compare the observed CMDs  between two Galactic models 
 using 2MASS data of the Galactic Bulge.
Clearly having  a realistic 3D extinction map is very 
important for such comparisons.  
 Both the Besan\c{c}on and TRILEGAL \citep{Girardi2005,Girardi2012} models of the 
 Galaxy  assume a simple, linear dust opacity,  $ \kappa~=~A_V/d$, 
 with $\kappa$ typically set to a constant  0.75\,mag/kpc \citep{Lynga1982}. 
As we already see in reality, the extinction-distance relation is not linear at all. 
As an example, we have applied
our 3D extinction map to Besan\c{c}on and TRILEGAL simulations
 of a field centred on $l~ = $
180$^{\circ}$ and $b=0^{\circ}$, and compared the results to 
the observed CMD from the XSTPS-GAC survey. 
A constant photometric error of 0.1\,mag is added to photometric magnitudes simulated by 
the Besan\c{c}on and TRILEGAL models.
A limiting magnitude, $r~<~$ 18.5\,mag, is also imposed.

\begin{figure}
\centering
   \includegraphics[width=0.42\textwidth]{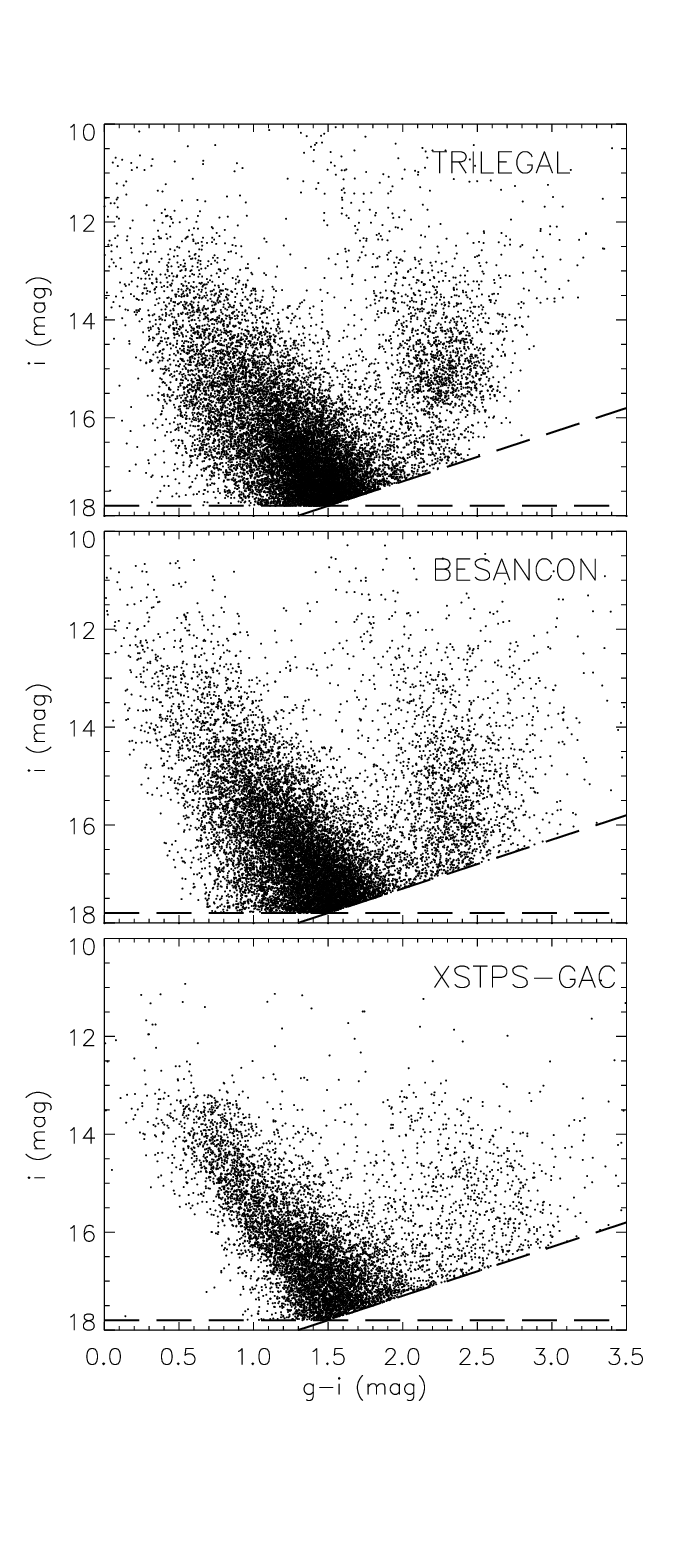}
\caption{Colour-magnitude diagrams for field centered on $l =180^{\circ}$, $b=0^{\circ}$, 
simulated by the  TRILEGAL (top panel) and
 Besan\c{c}on (middle) models, as well as produced by the observed data from 
XSTPS-GAC. The simulated CMDs have been reddened using the extinction-distance 
relation predicted by the 3D extinction map derived in the current work.
The dashed lines
delineate the limiting magnitude of completeness of the XSTPS-GAC
survey, estimated at $r=18.5$\,mag.} 
\label{cmdp}

\end{figure}

Fig.\,\ref{cmdp} shows that the CMDs produced by both models, 
reddened with extinction-distance relation 
from our 3D extinction map,
matches well the observed CMD from the XSTPS-GAC survey. 
We cut the observed and model data by the completeness
limits of the XSTPS-GAC survey, estimated at $r=18.5$\,mag. 
This completeness limit is indicated by dashed lines in Fig.\,\ref{cmdp}.
The CMD in the GAC region 
for $r~<~$18.5\,mag is dominated by two sequences: 
a sequence of dwarfs tilted towards the blue and 
a vertical sequence consisting mostly of the RC stars on the red. 
For the blue dwarf sequence, 
the stars become redder and redder as they go fainter and fainter 
(thus on average locate at larger and larger distances). 
We see a dramatical increase of reddening close to the limiting magnitude. 
We see an almost the same slope of the dwarf sequence for both  the simulated  and
observed data, indicating the robustness of 
 our extinction-distance relation. 
The simulated  dwarf sequence is more diffuse and less compact compared to  observed 
one, indicating that the 0.1\,mag of photometric error added to 
the model is probably overestimated, particularly for bright stars. 
For better comparison, a more realistic model of 
photometric errors for the XSTPS-GAC survey data 
(Yuan et al. 2014, in preparation) should be used instead.

In addition to the general trend of the CMD, 
the star number densities on the CMD 
predicted by the models are at large consistent with the observations, 
although both models seem to predict more stars than detected by the XSTPS-GAC.  
Finally, we note that the TRILEGAL model predicts a prominent RC population 
A clump structure at $i$ $\sim$ 14$^{\rm m}$.7 and $(g - i) \sim 2.35$\,mag. 
The feature is however less obviously in the simulated data of 
Besan\c{c}on as well as in the  XSTPS-GAC observed data.
\citet{Chen2013} notice that the Besan\c{c}on model predicts too high a
fraction of K\,giants at the Galactic center. In line with this, the current work also
finds that both the TRILEGAL and Besan\c{c}on models
seem to predict too much giants than actually observed.
 These findings should help improve the models.

\subsection{Comparison with CO maps}

\begin{figure*}
\centering
  \includegraphics[width=0.7\textwidth]{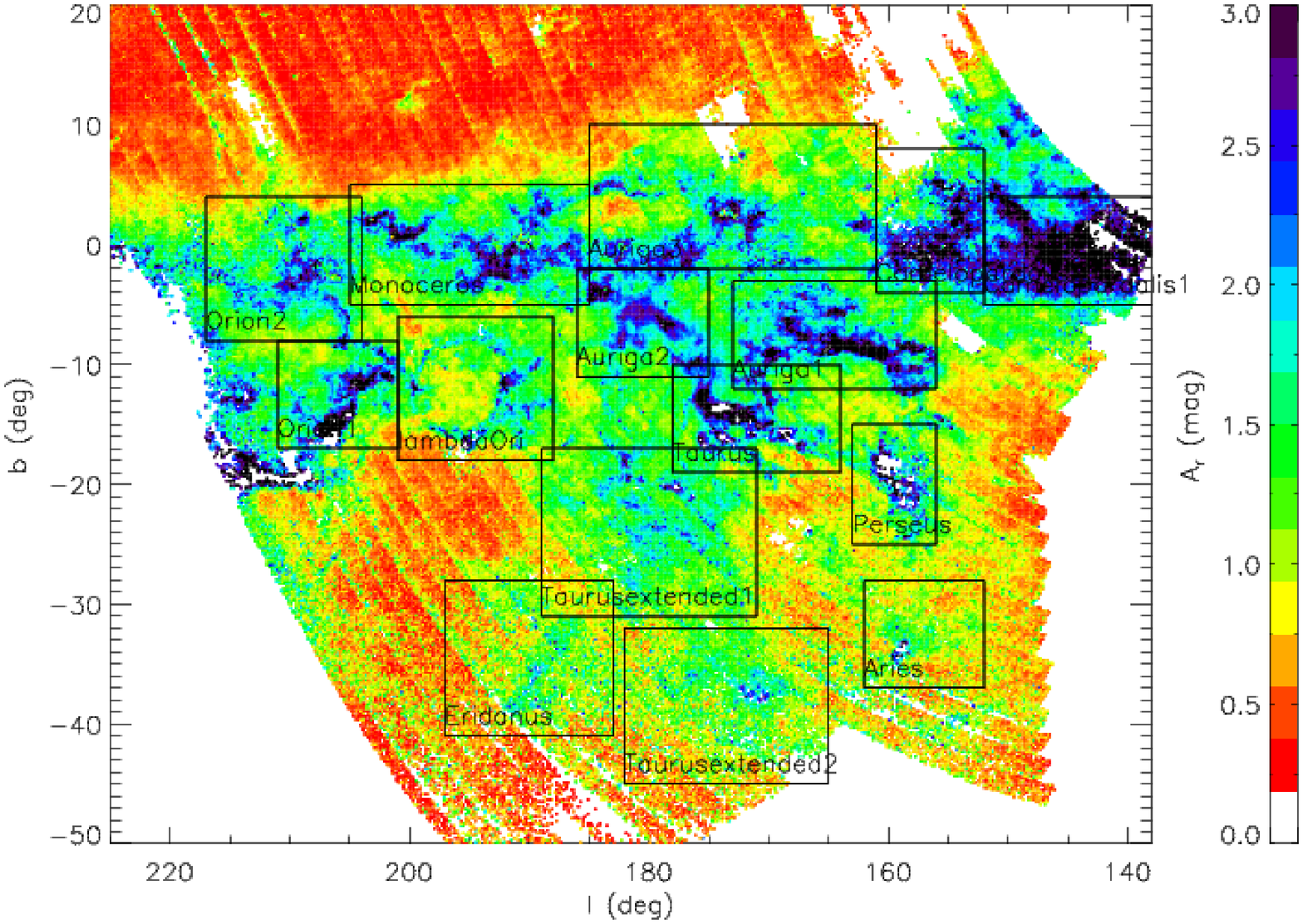} \\
  \includegraphics[width=0.7\textwidth]{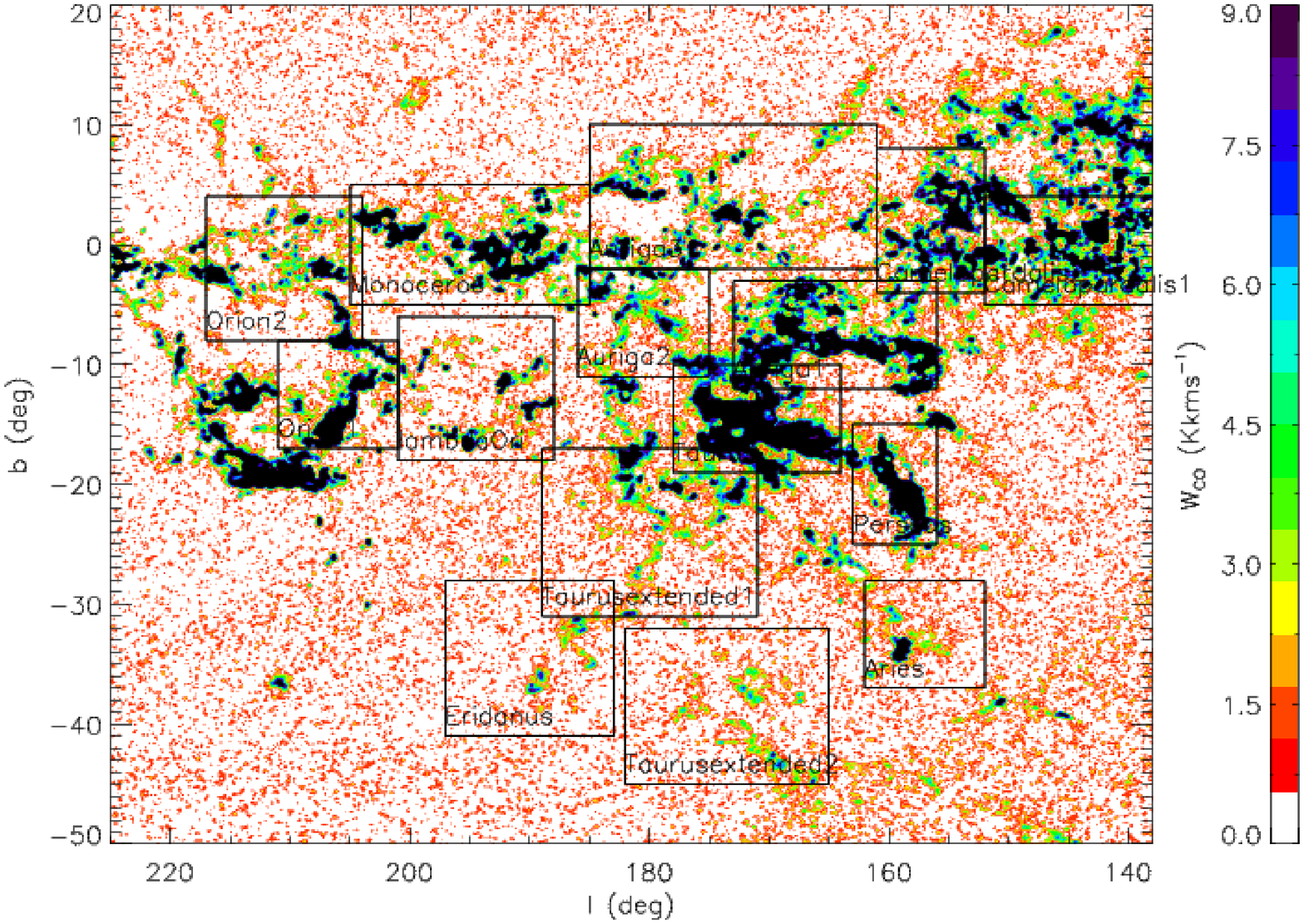}
\caption{The distributions of the extinction (upper panel) and
CO (lower panel) map in the GAC. The squares denote the
dust features mentioned in the text.}
\label{xfac}

\end{figure*}

Fig.\,\ref{xfac} compares our integrated dust extinction map  as given by  
the 90th percentile $A_r$ map of Sample\,A with   %$\rm ^{13}CO$ (J~=~0)
that of integrated CO map from the Plank survey \citep{Ade2013}.
The similarity of features revealed by the two maps is visually striking. 
The features already mentioned in Section\,4, including  the 
Orion region and the Taurus dark clouds, as well as some 
filamentary fine features are clearly visible on both maps.
Dust extinction at the optical and near-IR bands is dominated by large 
dust grains, while both common isotopologues of carbon monoxide, 
$\rm ^{12}$CO and $\rm ^{13}$CO, 
form in dusty environments, and their emission traces the gas content.
The features revealed by CO gas are thinned and sharper 
compared to dust features. This is because that
at the outer skirts of large molecular clouds where the  extinction is low, 
the CO is likely to be largely photodissociated,
producing a sharp distribution of CO emission, 
i.e. for regions of $A_r~<~1.0$\,mag, no CO emission is detected. Assuming a dust-to-gas ratio, 
 CO is often used as a proxy for 
dust extinction  (e.g., \citealt{Bok1977, Frerking1982, Langer1989, 
Dobashi2008, Liszt2014a, Liszt2014b}). A recent work on the dust-to-gas ratio, 
the so-called $X$ factor, is presented by \citet{Schultheis2013} 
for the region of  Galactic center. They obtain a 
$X~=~2.5 {\rm \times 10^{20} cm^{-2}\,K^{-1}\,km^{-1}\,s}$,
compatible to the value predicted by molecular cloud model of
\citet{Glover2011}.

Recently, the Planck Collaboration has released  CO emission maps 
of unprecedented quality \citep{Ade2013}. Those high quality CO maps, 
combined with the high quality dust maps obtained in the current work, should enable 
us to carry out a more detailed, quantitative study of
 the relationship between  dust extinction and CO emission. 
 Such a study will be presented in a future work.

\section{Conclusion}

By combing optical photometry from the XSTPS-GAC with those from 
2MASS and WISE in the near- and mid-IR, and using  a SED fitting algorithm 
similar to that employed by  \citet{Berry2011}, we have constructed  the first 
comprehensive quality 3D extinction map for a large sky area 
centered on the  GAC. The map  covers 6,000\,$\rm deg^2$ 
at angular resolutions between 3 -- 9\,arcmin, out to a distance of 4\,kpc. 
The extinction map is available online at CDS as well as 
at a website hosted at the Peking University 
(http://162.105.156.243/site/Photometric-Extinctions-and-Distances/). 
Compared to 2D SFD map, our 3D map should produce  better
results for disk stars of low latitudes, where the SFD map 
may have overestimated the extinction by a factor of 
2--3. A RJCE method is also used to generate a 2D  extinction map, 
yielding   results compatible  with those deduced from the best-fit SED algorithm. 
However for individual stars,
the RJCE method tends to yield results of larger uncertainty.
We show that by combining large multi-band surveys, such as the
XSTPS-GAC, 2MASS and WISE, one is able to constrain the interstellar extinction
for individual stars. The combination of optical and IR data can break 
the degeneracy between the effective temperature  $T_{\rm eff}$ 
[or equivalently the intrinsic colour $(g-i)_0$] and the
extinction $A_r$. By making use of more photometric bands that range from the optical to the IR, 
the SED fitting method yields results that are generally 
more robust  than the RJCE/NICE(R) method does, which
is based on two IR bands only.

In this process, we have also constructed  a reference catalog consisting 
of 132,316 stars of essentially zero extinction with quality photometry from the
XSTPS-GAC  for optical bands and from the  2MASS and WISE 
for IR bands. The reference catalog is used to construct standard 
stellar locus employed by the SED fitting algorithm. Both the reference 
catalog and the standard stellar locus are  available online from 
the aforementioned websites. 

The 3D extinction map presented in the current work should be quite useful for 
follow-up  analyse of LSS-GAC data, as well as data from other surveys. 

\section*{Acknowledgements}

We want to thank the referee for detailed and constructive comments 
that help improve
the manuscript significantly.
This work is partially supported by National Key Basic Research Program of China 
2014CB845700 and National Natural Science Foundation of China grant \#11078006.

This work has made use of data products from the Guoshoujing Telescope (the 
Large Sky Area Multi-Object Fibre Spectroscopic Telescope, LAMOST). LAMOST
is a National Major Scientific Project built by the Chinese Academy of 
Sciences. Funding for the project has been provided by the National
Development and Reform Commission. LAMOST is operated and managed by the 
National Astronomical Observatories, Chinese Academy of Sciences.

This publication makes use of data products from the Two Micron All Sky Survey, which is a joint
 project of the University of Massachusetts and the Infrared Processing and Analysis 
 Center/California Institute of Technology, funded by the National Aeronautics and Space 
 Administration and the National Science Foundation.

This publication makes use of data products from the Wide-field Infrared Survey Explorer, 
which is a joint project of the University of California, Los Angeles, and the Jet Propulsion
 Laboratory/California Institute of Technology, funded by the National Aeronautics and 
 Space Administration.

\bibliographystyle{mn2e} 
\bibliography{extinrev}

\begin{thebibliography}{102}
\expandafter\ifx\csname natexlab\endcsname\relax\def\natexlab#1{#1}\fi

\bibitem[{{An} {et~al.}(2008){An}, {Johnson}, {Clem}, {Yanny}, {Rockosi},
  {Morrison}, {Harding}, {Gunn}, {Allende Prieto}, {Beers}, {Cudworth},
  {Ivans}, {Ivezi{\'c}}, {Lee}, {Lupton}, {Bizyaev}, {Brewington},
  {Malanushenko}, {Malanushenko}, {Oravetz}, {Pan}, {Simmons}, {Snedden},
  {Watters}, \& {York}}]{An2008}
{An}, D., {et~al.} 2008, \apjs, 179, 326

\bibitem[{{Arce} \& {Goodman}(1999)}]{Arce1999}
{Arce}, H.~G. \& {Goodman}, A.~A. 1999, \apjl, 512, L135

\bibitem[{{Arenou} {et~al.}(1992){Arenou}, {Grenon}, \& {Gomez}}]{Arenou1992}
{Arenou}, F., {Grenon}, M., \& {Gomez}, A. 1992, \aap, 258, 104

\bibitem[{{Bailer-Jones}(2011)}]{Bailer-Jones2011}
{Bailer-Jones}, C.~A.~L. 2011, \mnras, 411, 435

\bibitem[{{Bajaja} {et~al.}(2005){Bajaja}, {Arnal}, {Larrarte}, {Morras},
  {P{\"o}ppel}, \& {Kalberla}}]{Bajaja2005}
{Bajaja}, E., {Arnal}, E.~M., {Larrarte}, J.~J., {Morras}, R., {P{\"o}ppel},
  W.~G.~L., \& {Kalberla}, P.~M.~W. 2005, \aap, 440, 767

\bibitem[{{Benjamin} {et~al.}(2005){Benjamin}, {Churchwell}, {Babler},
  {Indebetouw}, {Meade}, {Whitney}, {Watson}, {Wolfire}, {Wolff}, {Ignace},
  {Bania}, {Bracker}, {Clemens}, {Chomiuk}, {Cohen}, {Dickey}, {Jackson},
  {Kobulnicky}, {Mercer}, {Mathis}, {Stolovy}, \& {Uzpen}}]{Benjamin2005}
{Benjamin}, R.~A., {et~al.} 2005, \apjl, 630, L149

\bibitem[{{Berry} {et~al.}(2012){Berry}, {Ivezi{\'c}}, {Sesar}, {Juri{\'c}},
  {Schlafly}, {Bellovary}, {Finkbeiner}, {Vrbanec}, {Beers}, {Brooks},
  {Schneider}, {Gibson}, {Kimball}, {Jones}, {Yoachim}, {Krughoff}, {Connolly},
  {Loebman}, {Bond}, {Schlegel}, {Dalcanton}, {Yanny}, {Majewski}, {Knapp},
  {Gunn}, {Allyn Smith}, {Fukugita}, {Kent}, {Barentine}, {Krzesinski}, \&
  {Long}}]{Berry2011}
{Berry}, M., {et~al.} 2012, \apj, 757, 166

\bibitem[{{Boggess} {et~al.}(1992){Boggess}, {Mather}, {Weiss}, {Bennett},
  {Cheng}, {Dwek}, {Gulkis}, {Hauser}, {Janssen}, {Kelsall}, {Meyer},
  {Moseley}, {Murdock}, {Shafer}, {Silverberg}, {Smoot}, {Wilkinson}, \&
  {Wright}}]{Boggess1992}
{Boggess}, N.~W., {et~al.} 1992, \apj, 397, 420

\bibitem[{{Bok}(1977)}]{Bok1977}
{Bok}, B.~J. 1977, \pasp, 89, 597

\bibitem[{{Bond} {et~al.}(2010){Bond}, {Ivezi{\'c}}, {Sesar}, {Juri{\'c}},
  {Munn}, {Kowalski}, {Loebman}, {Ro{\v s}kar}, {Beers}, {Dalcanton},
  {Rockosi}, {Yanny}, {Newberg}, {Allende Prieto}, {Wilhelm}, {Lee},
  {Sivarani}, {Majewski}, {Norris}, {Bailer-Jones}, {Re Fiorentin}, {Schlegel},
  {Uomoto}, {Lupton}, {Knapp}, {Gunn}, {Covey}, {Allyn Smith}, {Miknaitis},
  {Doi}, {Tanaka}, {Fukugita}, {Kent}, {Finkbeiner}, {Quinn}, {Hawley},
  {Anderson}, {Kiuchi}, {Chen}, {Bushong}, {Sohi}, {Haggard}, {Kimball},
  {McGurk}, {Barentine}, {Brewington}, {Harvanek}, {Kleinman}, {Krzesinski},
  {Long}, {Nitta}, {Snedden}, {Lee}, {Pier}, {Harris}, {Brinkmann}, \&
  {Schneider}}]{Bond2010}
{Bond}, N.~A., {et~al.} 2010, \apj, 716, 1

\bibitem[{{Burstein} \& {Heiles}(1982)}]{Burstein1982}
{Burstein}, D. \& {Heiles}, C. 1982, \aj, 87, 1165

\bibitem[{{Cardelli} {et~al.}(1989){Cardelli}, {Clayton}, \&
  {Mathis}}]{Cardelli1989}
{Cardelli}, J.~A., {Clayton}, G.~C., \& {Mathis}, J.~S. 1989, \apj, 345, 245

\bibitem[{{Chen} {et~al.}(1999){Chen}, {Figueras}, {Torra}, {Jordi}, {Luri}, \&
  {Galad{\'{\i}}-Enr{\'{\i}}quez}}]{Chen1999}
{Chen}, B., {Figueras}, F., {Torra}, J., {Jordi}, C., {Luri}, X., \&
  {Galad{\'{\i}}-Enr{\'{\i}}quez}, D. 1999, \aap, 352, 459

\bibitem[{{Chen} {et~al.}(1998){Chen}, {Vergely}, {Valette}, \&
  {Carraro}}]{Chen1998}
{Chen}, B., {Vergely}, J.~L., {Valette}, B., \& {Carraro}, G. 1998, \aap, 336,
  137

\bibitem[{{Chen} {et~al.}(2013){Chen}, {Schultheis}, {Jiang}, {Gonzalez},
  {Robin}, {Rejkuba}, \& {Minniti}}]{Chen2013}
{Chen}, B.~Q., {Schultheis}, M., {Jiang}, B.~W., {Gonzalez}, O.~A., {Robin},
  A.~C., {Rejkuba}, M., \& {Minniti}, D. 2013, \aap, 550, A42

\bibitem[{{Covey} {et~al.}(2007){Covey}, {Ivezi{\'c}}, {Schlegel},
  {Finkbeiner}, {Padmanabhan}, {Lupton}, {Ag{\"u}eros}, {Bochanski}, {Hawley},
  {West}, {Seth}, {Kimball}, {Gogarten}, {Claire}, {Haggard}, {Kaib},
  {Schneider}, \& {Sesar}}]{Covey2007}
{Covey}, K.~R., {et~al.} 2007, \aj, 134, 2398

\bibitem[{{Dame} {et~al.}(2001){Dame}, {Hartmann}, \& {Thaddeus}}]{Dame2001}
{Dame}, T.~M., {Hartmann}, D., \& {Thaddeus}, P. 2001, \apj, 547, 792

\bibitem[{{Davenport} {et~al.}(2014){Davenport}, {Ivezic}, {Becker}, {Ruan},
  {Hunt-Walker}, {Covey}, {Lewis}, {AlSayyad}, \& {Anderson}}]{Davenport2014}
{Davenport}, J.~R.~A., {et~al.} 2014, ArXiv e-prints

\bibitem[{{Dobashi} {et~al.}(2008){Dobashi}, {Bernard}, {Hughes}, {Paradis},
  {Reach}, \& {Kawamura}}]{Dobashi2008}
{Dobashi}, K., {Bernard}, J.-P., {Hughes}, A., {Paradis}, D., {Reach}, W.~T.,
  \& {Kawamura}, A. 2008, \aap, 484, 205

\bibitem[{{Dobashi} {et~al.}(2005){Dobashi}, {Uehara}, {Kandori}, {Sakurai},
  {Kaiden}, {Umemoto}, \& {Sato}}]{Dobashi2005}
{Dobashi}, K., {Uehara}, H., {Kandori}, R., {Sakurai}, T., {Kaiden}, M.,
  {Umemoto}, T., \& {Sato}, F. 2005, \pasj, 57, 1

\bibitem[{{Drimmel} {et~al.}(2003){Drimmel}, {Cabrera-Lavers}, \&
  {L{\'o}pez-Corredoira}}]{Drimmel2003}
{Drimmel}, R., {Cabrera-Lavers}, A., \& {L{\'o}pez-Corredoira}, M. 2003, \aap,
  409, 205

\bibitem[{{Eisenstein} {et~al.}(2006){Eisenstein}, {Liebert}, {Harris},
  {Kleinman}, {Nitta}, {Silvestri}, {Anderson}, {Barentine}, {Brewington},
  {Brinkmann}, {Harvanek}, {Krzesi{\'n}ski}, {Neilsen}, {Long}, {Schneider}, \&
  {Snedden}}]{Eisenstein2006}
{Eisenstein}, D.~J., {et~al.} 2006, \apjs, 167, 40

\bibitem[{{Finlator} {et~al.}(2000){Finlator}, {Ivezi{\'c}}, {Fan}, {Strauss},
  {Knapp}, {Lupton}, {Gunn}, {Rockosi}, {Anderson}, {Csabai}, {Hennessy},
  {Hindsley}, {McKay}, {Nichol}, {Schneider}, {Smith}, {York}, \& {SDSS
  Collaboration}}]{Finlator2000}
{Finlator}, K., {et~al.} 2000, \aj, 120, 2615

\bibitem[{{Fitzpatrick}(1999)}]{Fitzpatrick1999}
{Fitzpatrick}, E.~L. 1999, \pasp, 111, 63

\bibitem[{{Fitzpatrick} \& {Massa}(2009)}]{Fitzpatrick2009}
{Fitzpatrick}, E.~L. \& {Massa}, D. 2009, \apj, 699, 1209

\bibitem[{{Flaherty} {et~al.}(2007){Flaherty}, {Pipher}, {Megeath}, {Winston},
  {Gutermuth}, {Muzerolle}, {Allen}, \& {Fazio}}]{Flaherty2007}
{Flaherty}, K.~M., {Pipher}, J.~L., {Megeath}, S.~T., {Winston}, E.~M.,
  {Gutermuth}, R.~A., {Muzerolle}, J., {Allen}, L.~E., \& {Fazio}, G.~G. 2007,
  \apj, 663, 1069

\bibitem[{{Frerking} {et~al.}(1982){Frerking}, {Langer}, \&
  {Wilson}}]{Frerking1982}
{Frerking}, M.~A., {Langer}, W.~D., \& {Wilson}, R.~W. 1982, \apj, 262, 590

\bibitem[{{Froebrich} {et~al.}(2007){Froebrich}, {Murphy}, {Smith}, {Walsh}, \&
  {Del Burgo}}]{Froebrich2007}
{Froebrich}, D., {Murphy}, G.~C., {Smith}, M.~D., {Walsh}, J., \& {Del Burgo},
  C. 2007, \mnras, 378, 1447

\bibitem[{{Gao} {et~al.}(2009){Gao}, {Jiang}, \& {Li}}]{Gao2009}
{Gao}, J., {Jiang}, B.~W., \& {Li}, A. 2009, \apj, 707, 89

\bibitem[{{Geller} {et~al.}(2010){Geller}, {Mathieu}, {Braden}, {Meibom},
  {Platais}, \& {Dolan}}]{Geller2010}
{Geller}, A.~M., {Mathieu}, R.~D., {Braden}, E.~K., {Meibom}, S., {Platais},
  I., \& {Dolan}, C.~J. 2010, \aj, 139, 1383

\bibitem[{{Girardi} {et~al.}(2012){Girardi}, {Barbieri}, {Groenewegen},
  {Marigo}, {Bressan}, {Rocha-Pinto}, {Santiago}, {Camargo}, \& {da
  Costa}}]{Girardi2012}
{Girardi}, L., {et~al.} 2012, {TRILEGAL, a TRIdimensional modeL of thE GALaxy:
  Status and Future}, ed. A.~{Miglio}, J.~{Montalb{\'a}n}, \& A.~{Noels}, 165

\bibitem[{{Girardi} {et~al.}(2002){Girardi}, {Bertelli}, {Bressan}, {Chiosi},
  {Groenewegen}, {Marigo}, {Salasnich}, \& {Weiss}}]{Girardi2002}
{Girardi}, L., {Bertelli}, G., {Bressan}, A., {Chiosi}, C., {Groenewegen},
  M.~A.~T., {Marigo}, P., {Salasnich}, B., \& {Weiss}, A. 2002, \aap, 391, 195

\bibitem[{{Girardi} {et~al.}(2005){Girardi}, {Groenewegen}, {Hatziminaoglou},
  \& {da Costa}}]{Girardi2005}
{Girardi}, L., {Groenewegen}, M.~A.~T., {Hatziminaoglou}, E., \& {da Costa}, L.
  2005, \aap, 436, 895

\bibitem[{{Glover} \& {Mac Low}(2011)}]{Glover2011}
{Glover}, S.~C.~O. \& {Mac Low}, M.-M. 2011, \mnras, 412, 337

\bibitem[{{Gonzalez} {et~al.}(2012){Gonzalez}, {Rejkuba}, {Zoccali}, {Valenti},
  {Minniti}, {Schultheis}, {Tobar}, \& {Chen}}]{Gonzalez2012}
{Gonzalez}, O.~A., {Rejkuba}, M., {Zoccali}, M., {Valenti}, E., {Minniti}, D.,
  {Schultheis}, M., {Tobar}, R., \& {Chen}, B. 2012, \aap, 543, A13

\bibitem[{{Green} {et~al.}(2014){Green}, {Schlafly}, {Finkbeiner}, {Juri{\'c}},
  {Rix}, {Burgett}, {Chambers}, {Draper}, {Flewelling}, {Kudritzki}, {Magnier},
  {Martin}, {Metcalfe}, {Tonry}, {Wainscoat}, \& {Waters}}]{Green2014}
{Green}, G.~M., {et~al.} 2014, \apj, 783, 114

\bibitem[{{Hanson} \& {Bailer-Jones}(2014)}]{Hanson2013}
{Hanson}, R.~J. \& {Bailer-Jones}, C.~A.~L. 2014, \mnras, 438, 2938

\bibitem[{{Hawley} {et~al.}(2002){Hawley}, {Covey}, {Knapp}, {Golimowski},
  {Fan}, {Anderson}, {Gunn}, {Harris}, {Ivezi{\'c}}, {Long}, {Lupton},
  {McGehee}, {Narayanan}, {Peng}, {Schlegel}, {Schneider}, {Spahn}, {Strauss},
  {Szkody}, {Tsvetanov}, {Walkowicz}, {Brinkmann}, {Harvanek}, {Hennessy},
  {Kleinman}, {Krzesinski}, {Long}, {Neilsen}, {Newman}, {Nitta}, {Snedden}, \&
  {York}}]{Hawley2002}
{Hawley}, S.~L., {et~al.} 2002, \aj, 123, 3409

\bibitem[{{Heiles}(1998)}]{Heiles1998}
{Heiles}, C. 1998, \apj, 498, 689

\bibitem[{{High} {et~al.}(2009){High}, {Stubbs}, {Rest}, {Stalder}, \&
  {Challis}}]{High2009}
{High}, F.~W., {Stubbs}, C.~W., {Rest}, A., {Stalder}, B., \& {Challis}, P.
  2009, \aj, 138, 110

\bibitem[{{Indebetouw} {et~al.}(2005){Indebetouw}, {Mathis}, {Babler}, {Meade},
  {Watson}, {Whitney}, {Wolff}, {Wolfire}, {Cohen}, {Bania}, {Benjamin},
  {Clemens}, {Dickey}, {Jackson}, {Kobulnicky}, {Marston}, {Mercer},
  {Stauffer}, {Stolovy}, \& {Churchwell}}]{Indebetouw2005}
{Indebetouw}, R., {et~al.} 2005, \apj, 619, 931

\bibitem[{{Ivezi{\'c}} {et~al.}(2008){Ivezi{\'c}}, {Sesar}, {Juri{\'c}},
  {Bond}, {Dalcanton}, {Rockosi}, {Yanny}, {Newberg}, {Beers}, {Allende
  Prieto}, {Wilhelm}, {Lee}, {Sivarani}, {Norris}, {Bailer-Jones}, {Re
  Fiorentin}, {Schlegel}, {Uomoto}, {Lupton}, {Knapp}, {Gunn}, {Covey},
  {Smith}, {Miknaitis}, {Doi}, {Tanaka}, {Fukugita}, {Kent}, {Finkbeiner},
  {Munn}, {Pier}, {Quinn}, {Hawley}, {Anderson}, {Kiuchi}, {Chen}, {Bushong},
  {Sohi}, {Haggard}, {Kimball}, {Barentine}, {Brewington}, {Harvanek},
  {Kleinman}, {Krzesinski}, {Long}, {Nitta}, {Snedden}, {Lee}, {Harris},
  {Brinkmann}, {Schneider}, \& {York}}]{Ivezic2008}
{Ivezi{\'c}}, {\v Z}., {et~al.} 2008, \apj, 684, 287

\bibitem[{{Ivezi{\'c}} {et~al.}(2007){Ivezi{\'c}}, {Smith}, {Miknaitis}, {Lin},
  {Tucker}, {Lupton}, {Gunn}, {Knapp}, {Strauss}, {Sesar}, {Doi}, {Tanaka},
  {Fukugita}, {Holtzman}, {Kent}, {Yanny}, {Schlegel}, {Finkbeiner},
  {Padmanabhan}, {Rockosi}, {Juri{\'c}}, {Bond}, {Lee}, {Stoughton}, {Jester},
  {Harris}, {Harding}, {Morrison}, {Brinkmann}, {Schneider}, \&
  {York}}]{Ivezic2007}
{Ivezi{\'c}}, {\v Z}., {et~al.} 2007, \aj, 134, 973

\bibitem[{{Jiang} {et~al.}(2006){Jiang}, {Gao}, {Omont}, {Schuller}, \&
  {Simon}}]{Jiang2006}
{Jiang}, B.~W., {Gao}, J., {Omont}, A., {Schuller}, F., \& {Simon}, G. 2006,
  \aap, 446, 551

\bibitem[{{Jones} {et~al.}(2011){Jones}, {West}, \& {Foster}}]{Jones2011}
{Jones}, D.~O., {West}, A.~A., \& {Foster}, J.~B. 2011, \aj, 142, 44

\bibitem[{{Juri{\'c}} {et~al.}(2008){Juri{\'c}}, {Ivezi{\'c}}, {Brooks},
  {Lupton}, {Schlegel}, {Finkbeiner}, {Padmanabhan}, {Bond}, {Sesar},
  {Rockosi}, {Knapp}, {Gunn}, {Sumi}, {Schneider}, {Barentine}, {Brewington},
  {Brinkmann}, {Fukugita}, {Harvanek}, {Kleinman}, {Krzesinski}, {Long},
  {Neilsen}, {Nitta}, {Snedden}, \& {York}}]{Juric2008}
{Juri{\'c}}, M., {et~al.} 2008, \apj, 673, 864

\bibitem[{{Kalberla} {et~al.}(2005){Kalberla}, {Burton}, {Hartmann}, {Arnal},
  {Bajaja}, {Morras}, \& {P{\"o}ppel}}]{Kalberla2005}
{Kalberla}, P.~M.~W., {Burton}, W.~B., {Hartmann}, D., {Arnal}, E.~M.,
  {Bajaja}, E., {Morras}, R., \& {P{\"o}ppel}, W.~G.~L. 2005, \aap, 440, 775

\bibitem[{{Kalirai} {et~al.}(2003){Kalirai}, {Fahlman}, {Richer}, \&
  {Ventura}}]{Kalirai2003}
{Kalirai}, J.~S., {Fahlman}, G.~G., {Richer}, H.~B., \& {Ventura}, P. 2003,
  \aj, 126, 1402

\bibitem[{{Lallement} {et~al.}(2014){Lallement}, {Vergely}, {Valette},
  {Puspitarini}, {Eyer}, \& {Casagrande}}]{Lallement2013}
{Lallement}, R., {Vergely}, J.-L., {Valette}, B., {Puspitarini}, L., {Eyer},
  L., \& {Casagrande}, L. 2014, \aap, 561, A91

\bibitem[{{Langer} {et~al.}(1989){Langer}, {Wilson}, {Goldsmith}, \&
  {Beichman}}]{Langer1989}
{Langer}, W.~D., {Wilson}, R.~W., {Goldsmith}, P.~F., \& {Beichman}, C.~A.
  1989, \apj, 337, 355

\bibitem[{{Lejeune} {et~al.}(1997){Lejeune}, {Cuisinier}, \&
  {Buser}}]{Lejeune1997}
{Lejeune}, T., {Cuisinier}, F., \& {Buser}, R. 1997, \aaps, 125, 229

\bibitem[{{Lejeune} {et~al.}(1998){Lejeune}, {Cuisinier}, \&
  {Buser}}]{Lejeune1998}
{Lejeune}, T., {Cuisinier}, F., \& {Buser}, R. 1998, \aaps, 130, 65

\bibitem[{{Liszt}(2014{\natexlab{a}})}]{Liszt2014a}
{Liszt}, H. 2014{\natexlab{a}}, \apj, 783, 17

\bibitem[{{Liszt}(2014{\natexlab{b}})}]{Liszt2014b}
{Liszt}, H. 2014{\natexlab{b}}, \apj, 780, 10

\bibitem[{Liu {et~al.}(2014)Liu, Yuan, Huo, Deng, Hou, Zhao, Zhao, Shi, Luo,
  Xiang, Zhang, Huang, \& Zhang}]{Liu2013}
Liu, X.-W., {et~al.} 2014, in Feltzing S., Zhao G., Walton N., Whitelock P.,
  eds, Proc. IAU Symp. 298, Setting the scene for Gaia and LAMOST, Cambridge
  University Press, pp. 310-321, preprint (arXiv: 1306.5376)

\bibitem[{{Lombardi} \& {Alves}(2001)}]{Lombardi2001a}
{Lombardi}, M. \& {Alves}, J. 2001, \aap, 377, 1023

\bibitem[{{Lombardi} {et~al.}(2011){Lombardi}, {Alves}, \&
  {Lada}}]{Lombardi2011b}
{Lombardi}, M., {Alves}, J., \& {Lada}, C.~J. 2011, \aap, 535, A16

\bibitem[{{Lynga}(1982)}]{Lynga1982}
{Lynga}, G. 1982, \aap, 109, 213

\bibitem[{{Majewski} {et~al.}(2011){Majewski}, {Zasowski}, \&
  {Nidever}}]{Majewski2011}
{Majewski}, S.~R., {Zasowski}, G., \& {Nidever}, D.~L. 2011, \apj, 739, 25

\bibitem[{{Marshall} {et~al.}(2009){Marshall}, {Joncas}, \&
  {Jones}}]{Marshall2009}
{Marshall}, D.~J., {Joncas}, G., \& {Jones}, A.~P. 2009, \apj, 706, 727

\bibitem[{{Marshall} {et~al.}(2006){Marshall}, {Robin}, {Reyl{\'e}},
  {Schultheis}, \& {Picaud}}]{Marshall2006}
{Marshall}, D.~J., {Robin}, A.~C., {Reyl{\'e}}, C., {Schultheis}, M., \&
  {Picaud}, S. 2006, \aap, 453, 635

\bibitem[{{Martin} {et~al.}(2005){Martin}, {Fanson}, {Schiminovich},
  {Morrissey}, {Friedman}, {Barlow}, {Conrow}, {Grange}, {Jelinsky},
  {Milliard}, {Siegmund}, {Bianchi}, {Byun}, {Donas}, {Forster}, {Heckman},
  {Lee}, {Madore}, {Malina}, {Neff}, {Rich}, {Small}, {Surber}, {Szalay},
  {Welsh}, \& {Wyder}}]{Martin2005}
{Martin}, D.~C., {et~al.} 2005, \apjl, 619, L1

\bibitem[{{Mathis} \& {Cardelli}(1990)}]{Mathis1990}
{Mathis}, J.~S. \& {Cardelli}, J.~A. 1990, in Bulletin of the American
  Astronomical Society, Vol.~22, Bulletin of the American Astronomical Society,
  861

\bibitem[{{Meisner} \& {Finkbeiner}(2014)}]{Meisner2013}
{Meisner}, A.~M. \& {Finkbeiner}, D.~P. 2014, \apj, 781, 5

\bibitem[{{Morrison} {et~al.}(2001){Morrison}, {McLean}, \& {GSC-Catalog II
  Construction Team}}]{Morrison2001}
{Morrison}, J.~E., {McLean}, B., \& {GSC-Catalog II Construction Team}. 2001,
  in Bulletin of the American Astronomical Society, Vol.~33, AAS/Division of
  Dynamical Astronomy Meeting \#32, 1194

\bibitem[{{Neckel} {et~al.}(1980){Neckel}, {Klare}, \&
  {Sarcander}}]{Neckel1980}
{Neckel}, T., {Klare}, G., \& {Sarcander}, M. 1980, \aaps, 42, 251

\bibitem[{{Neugebauer} {et~al.}(1984){Neugebauer}, {Habing}, {van Duinen},
  {Aumann}, {Baud}, {Beichman}, {Beintema}, {Boggess}, {Clegg}, {de Jong},
  {Emerson}, {Gautier}, {Gillett}, {Harris}, {Hauser}, {Houck}, {Jennings},
  {Low}, {Marsden}, {Miley}, {Olnon}, {Pottasch}, {Raimond}, {Rowan-Robinson},
  {Soifer}, {Walker}, {Wesselius}, \& {Young}}]{Neugebauer1984}
{Neugebauer}, G., {et~al.} 1984, \apjl, 278, L1

\bibitem[{{Nidever} {et~al.}(2012){Nidever}, {Zasowski}, \&
  {Majewski}}]{Nidever2012}
{Nidever}, D.~L., {Zasowski}, G., \& {Majewski}, S.~R. 2012, \apjs, 201, 35

\bibitem[{{Nishiyama} {et~al.}(2009){Nishiyama}, {Tamura}, {Hatano}, {Kato},
  {Tanab{\'e}}, {Sugitani}, \& {Nagata}}]{Nishiyama2009}
{Nishiyama}, S., {Tamura}, M., {Hatano}, H., {Kato}, D., {Tanab{\'e}}, T.,
  {Sugitani}, K., \& {Nagata}, T. 2009, \apj, 696, 1407

\bibitem[{{O'Donnell}(1994)}]{ODonnell1994}
{O'Donnell}, J.~E. 1994, \apj, 422, 158

\bibitem[{{Peek} \& {Graves}(2010)}]{Peek2010}
{Peek}, J.~E.~G. \& {Graves}, G.~J. 2010, \apj, 719, 415

\bibitem[{{Planck Collaboration} {et~al.}(2013{\natexlab{a}}){Planck
  Collaboration}, {Abergel}, {Ade}, {Aghanim}, {Alina}, {Alves},
  {Armitage-Caplan}, {Arnaud}, {Ashdown}, {Atrio-Barandela}, \&
  et~al.}]{Abergel2013}
{Planck Collaboration}, {et~al.} 2013{\natexlab{a}}, ArXiv e-prints

\bibitem[{{Planck Collaboration} {et~al.}(2013{\natexlab{b}}){Planck
  Collaboration}, {Ade}, {Aghanim}, {Alves}, {Armitage-Caplan}, {Arnaud},
  {Ashdown}, {Atrio-Barandela}, {Aumont}, {Baccigaluppi}, \& et~al.}]{Ade2013}
{Planck Collaboration}, {et~al.} 2013{\natexlab{b}}, ArXiv e-prints

\bibitem[{{Randich} {et~al.}(2006){Randich}, {Sestito}, {Primas},
  {Pallavicini}, \& {Pasquini}}]{Randich2006}
{Randich}, S., {Sestito}, P., {Primas}, F., {Pallavicini}, R., \& {Pasquini},
  L. 2006, \aap, 450, 557

\bibitem[{{Richards} {et~al.}(2001){Richards}, {Fan}, {Schneider}, {Vanden
  Berk}, {Strauss}, {York}, {Anderson}, {Anderson}, {Annis}, {Bahcall},
  {Bernardi}, {Briggs}, {Brinkmann}, {Brunner}, {Burles}, {Carey}, {Castander},
  {Connolly}, {Crocker}, {Csabai}, {Doi}, {Finkbeiner}, {Friedman}, {Frieman},
  {Fukugita}, {Gunn}, {Hindsley}, {Ivezi{\'c}}, {Kent}, {Knapp}, {Lamb},
  {Leger}, {Long}, {Loveday}, {Lupton}, {McKay}, {Meiksin}, {Merrelli}, {Munn},
  {Newberg}, {Newcomb}, {Nichol}, {Owen}, {Pier}, {Pope}, {Richmond},
  {Rockosi}, {Schlegel}, {Siegmund}, {Smee}, {Snir}, {Stoughton}, {Stubbs},
  {SubbaRao}, {Szalay}, {Szokoly}, {Tremonti}, {Uomoto}, {Waddell}, {Yanny}, \&
  {Zheng}}]{Richards2001}
{Richards}, G.~T., {et~al.} 2001, \aj, 121, 2308

\bibitem[{{Robin} {et~al.}(2012){Robin}, {Marshall}, {Schultheis}, \&
  {Reyl{\'e}}}]{Robin2012}
{Robin}, A.~C., {Marshall}, D.~J., {Schultheis}, M., \& {Reyl{\'e}}, C. 2012,
  \aap, 538, A106

\bibitem[{{Robin} {et~al.}(2003){Robin}, {Reyl{\'e}}, {Derri{\`e}re}, \&
  {Picaud}}]{Robin2003}
{Robin}, A.~C., {Reyl{\'e}}, C., {Derri{\`e}re}, S., \& {Picaud}, S. 2003,
  \aap, 409, 523

\bibitem[{{Rowles} \& {Froebrich}(2009)}]{Rowles2009}
{Rowles}, J. \& {Froebrich}, D. 2009, \mnras, 395, 1640

\bibitem[{{Sale}(2012)}]{Sale2012}
{Sale}, S.~E. 2012, \mnras, 427, 2119

\bibitem[{{Sale} {et~al.}(2009){Sale}, {Drew}, {Unruh}, {Irwin}, {Knigge},
  {Phillipps}, {Zijlstra}, {G{\"a}nsicke}, {Greimel}, {Groot}, {Mampaso},
  {Morris}, {Napiwotzki}, {Steeghs}, \& {Walton}}]{Sale2009}
{Sale}, S.~E., {et~al.} 2009, \mnras, 392, 497

\bibitem[{{Schlafly} \& {Finkbeiner}(2011)}]{Schlafly2011}
{Schlafly}, E.~F. \& {Finkbeiner}, D.~P. 2011, \apj, 737, 103

\bibitem[{{Schlafly} {et~al.}(2010){Schlafly}, {Finkbeiner}, {Schlegel},
  {Juri{\'c}}, {Ivezi{\'c}}, {Gibson}, {Knapp}, \& {Weaver}}]{Schlafly2010}
{Schlafly}, E.~F., {Finkbeiner}, D.~P., {Schlegel}, D.~J., {Juri{\'c}}, M.,
  {Ivezi{\'c}}, {\v Z}., {Gibson}, R.~R., {Knapp}, G.~R., \& {Weaver}, B.~A.
  2010, \apj, 725, 1175

\bibitem[{{Schlafly} {et~al.}(2014){Schlafly}, {Green}, {Finkbeiner}, {Rix},
  {Bell}, {Burgett}, {Chambers}, {Draper}, {Hodapp}, {Kaiser}, {Magnier},
  {Martin}, {Metcalfe}, {Price}, \& {Tonry}}]{Schlafly2014}
{Schlafly}, E.~F., {et~al.} 2014, \apj, 786, 29

\bibitem[{{Schlegel} {et~al.}(1998){Schlegel}, {Finkbeiner}, \&
  {Davis}}]{Schlegel1998}
{Schlegel}, D.~J., {Finkbeiner}, D.~P., \& {Davis}, M. 1998, \apj, 500, 525

\bibitem[{{Schultheis} {et~al.}(2014){Schultheis}, {Chen}, {Jiang}, {Gonzalez},
  {Enokiya}, {Fukui}, {Torii}, {Rejkuba}, \& {Minniti}}]{Schultheis2013}
{Schultheis}, M., {et~al.} 2014, ArXiv e-prints

\bibitem[{{Schultheis} {et~al.}(1999){Schultheis}, {Ganesh}, {Simon}, {Omont},
  {Alard}, {Borsenberger}, {Copet}, {Epchtein}, {Fouqu{\'e}}, \&
  {Habing}}]{Schultheis1999}
{Schultheis}, M., {et~al.} 1999, \aap, 349, L69

\bibitem[{{Skrutskie} {et~al.}(1997){Skrutskie}, {Schneider}, {Stiening},
  {Strom}, {Weinberg}, {Beichman}, {Chester}, {Cutri}, {Lonsdale}, {Elias},
  {Elston}, {Capps}, {Carpenter}, {Huchra}, {Liebert}, {Monet}, {Price}, \&
  {Seitzer}}]{Skrutskie1997}
{Skrutskie}, M.~F., {et~al.} 1997, in Astrophysics and Space Science Library,
  Vol. 210, The Impact of Large Scale Near-IR Sky Surveys, ed. F.~{Garzon},
  N.~{Epchtein}, A.~{Omont}, B.~{Burton}, \& P.~{Persi}, 25

\bibitem[{{Smol{\v c}i{\'c}} {et~al.}(2004){Smol{\v c}i{\'c}}, {Ivezi{\'c}},
  {Knapp}, {Lupton}, {Pavlovski}, {Iliji{\'c}}, {Schlegel}, {Smith}, {McGehee},
  {Silvestri}, {Hawley}, {Rockosi}, {Gunn}, {Strauss}, {Fan}, {Eisenstein}, \&
  {Harris}}]{Smolcic2004}
{Smol{\v c}i{\'c}}, V., {et~al.} 2004, \apjl, 615, L141

\bibitem[{{Strai{\v z}ys} \& {Laugalys}(2008)}]{Straizys2008}
{Strai{\v z}ys}, V. \& {Laugalys}, V. 2008, Baltic Astronomy, 17, 1

\bibitem[{{Straizys} {et~al.}(1992){Straizys}, {Cernis}, {Kazlauskas}, \&
  {Meistas}}]{Straizys1992}
{Straizys}, V., {Cernis}, K., {Kazlauskas}, A., \& {Meistas}, E. 1992, Baltic
  Astronomy, 1, 149

\bibitem[{{Taylor} {et~al.}(2008){Taylor}, {Joner}, \& {Jeffery}}]{Taylor2008}
{Taylor}, B.~J., {Joner}, M.~D., \& {Jeffery}, E.~J. 2008, \apjs, 176, 262

\bibitem[{{Uttenthaler} {et~al.}(2012){Uttenthaler}, {Schultheis}, {Nataf},
  {Robin}, {Lebzelter}, \& {Chen}}]{Uttenthaler2012}
{Uttenthaler}, S., {Schultheis}, M., {Nataf}, D.~M., {Robin}, A.~C.,
  {Lebzelter}, T., \& {Chen}, B. 2012, \aap, 546, A57

\bibitem[{{Wright} {et~al.}(2010){Wright}, {Eisenhardt}, {Mainzer}, {Ressler},
  {Cutri}, {Jarrett}, {Kirkpatrick}, {Padgett}, {McMillan}, {Skrutskie},
  {Stanford}, {Cohen}, {Walker}, {Mather}, {Leisawitz}, {Gautier}, {McLean},
  {Benford}, {Lonsdale}, {Blain}, {Mendez}, {Irace}, {Duval}, {Liu}, {Royer},
  {Heinrichsen}, {Howard}, {Shannon}, {Kendall}, {Walsh}, {Larsen}, {Cardon},
  {Schick}, {Schwalm}, {Abid}, {Fabinsky}, {Naes}, \& {Tsai}}]{Wright2010}
{Wright}, E.~L., {et~al.} 2010, \aj, 140, 1868

\bibitem[{{Wu} {et~al.}(2009){Wu}, {Zhou}, {Ma}, \& {Du}}]{Wu2009}
{Wu}, Z.-Y., {Zhou}, X., {Ma}, J., \& {Du}, C.-H. 2009, \mnras, 399, 2146

\bibitem[{{Xin} \& {Deng}(2005)}]{Xin2005}
{Xin}, Y. \& {Deng}, L. 2005, \apj, 619, 824

\bibitem[{{Yadav} {et~al.}(2008){Yadav}, {Bedin}, {Piotto}, {Anderson},
  {Cassisi}, {Villanova}, {Platais}, {Pasquini}, {Momany}, \&
  {Sagar}}]{Yadav2008}
{Yadav}, R.~K.~S., {et~al.} 2008, \aap, 484, 609

\bibitem[{{Yanny} {et~al.}(2009){Yanny}, {Rockosi}, {Newberg}, {Knapp},
  {Adelman-McCarthy}, {Alcorn}, {Allam}, {Allende Prieto}, {An}, {Anderson},
  {Anderson}, {Bailer-Jones}, {Bastian}, {Beers}, {Bell}, {Belokurov},
  {Bizyaev}, {Blythe}, {Bochanski}, {Boroski}, {Brinchmann}, {Brinkmann},
  {Brewington}, {Carey}, {Cudworth}, {Evans}, {Evans}, {Gates}, {G{\"a}nsicke},
  {Gillespie}, {Gilmore}, {Nebot Gomez-Moran}, {Grebel}, {Greenwell}, {Gunn},
  {Jordan}, {Jordan}, {Harding}, {Harris}, {Hendry}, {Holder}, {Ivans},
  {Ivezi{\v c}}, {Jester}, {Johnson}, {Kent}, {Kleinman}, {Kniazev},
  {Krzesinski}, {Kron}, {Kuropatkin}, {Lebedeva}, {Lee}, {French Leger},
  {L{\'e}pine}, {Levine}, {Lin}, {Long}, {Loomis}, {Lupton}, {Malanushenko},
  {Malanushenko}, {Margon}, {Martinez-Delgado}, {McGehee}, {Monet}, {Morrison},
  {Munn}, {Neilsen}, {Nitta}, {Norris}, {Oravetz}, {Owen}, {Padmanabhan},
  {Pan}, {Peterson}, {Pier}, {Platson}, {Re Fiorentin}, {Richards}, {Rix},
  {Schlegel}, {Schneider}, {Schreiber}, {Schwope}, {Sibley}, {Simmons},
  {Snedden}, {Allyn Smith}, {Stark}, {Stauffer}, {Steinmetz}, {Stoughton},
  {SubbaRao}, {Szalay}, {Szkody}, {Thakar}, {Sivarani}, {Tucker}, {Uomoto},
  {Vanden Berk}, {Vidrih}, {Wadadekar}, {Watters}, {Wilhelm}, {Wyse}, {Yarger},
  \& {Zucker}}]{Yanny2009}
{Yanny}, B., {et~al.} 2009, \aj, 137, 4377

\bibitem[{{York} {et~al.}(2000){York}, {Adelman}, {Anderson}, {Anderson},
  {Annis}, {Bahcall}, {Bakken}, {Barkhouser}, {Bastian}, {Berman}, {Boroski},
  {Bracker}, {Briegel}, {Briggs}, {Brinkmann}, {Brunner}, {Burles}, {Carey},
  {Carr}, {Castander}, {Chen}, {Colestock}, {Connolly}, {Crocker}, {Csabai},
  {Czarapata}, {Davis}, {Doi}, {Dombeck}, {Eisenstein}, {Ellman}, {Elms},
  {Evans}, {Fan}, {Federwitz}, {Fiscelli}, {Friedman}, {Frieman}, {Fukugita},
  {Gillespie}, {Gunn}, {Gurbani}, {de Haas}, {Haldeman}, {Harris}, {Hayes},
  {Heckman}, {Hennessy}, {Hindsley}, {Holm}, {Holmgren}, {Huang}, {Hull},
  {Husby}, {Ichikawa}, {Ichikawa}, {Ivezi{\'c}}, {Kent}, {Kim}, {Kinney},
  {Klaene}, {Kleinman}, {Kleinman}, {Knapp}, {Korienek}, {Kron}, {Kunszt},
  {Lamb}, {Lee}, {Leger}, {Limmongkol}, {Lindenmeyer}, {Long}, {Loomis},
  {Loveday}, {Lucinio}, {Lupton}, {MacKinnon}, {Mannery}, {Mantsch}, {Margon},
  {McGehee}, {McKay}, {Meiksin}, {Merelli}, {Monet}, {Munn}, {Narayanan},
  {Nash}, {Neilsen}, {Neswold}, {Newberg}, {Nichol}, {Nicinski}, {Nonino},
  {Okada}, {Okamura}, {Ostriker}, {Owen}, {Pauls}, {Peoples}, {Peterson},
  {Petravick}, {Pier}, {Pope}, {Pordes}, {Prosapio}, {Rechenmacher}, {Quinn},
  {Richards}, {Richmond}, {Rivetta}, {Rockosi}, {Ruthmansdorfer}, {Sandford},
  {Schlegel}, {Schneider}, {Sekiguchi}, {Sergey}, {Shimasaku}, {Siegmund},
  {Smee}, {Smith}, {Snedden}, {Stone}, {Stoughton}, {Strauss}, {Stubbs},
  {SubbaRao}, {Szalay}, {Szapudi}, {Szokoly}, {Thakar}, {Tremonti}, {Tucker},
  {Uomoto}, {Vanden Berk}, {Vogeley}, {Waddell}, {Wang}, {Watanabe},
  {Weinberg}, {Yanny}, {Yasuda}, \& {SDSS Collaboration}}]{York2000}
{York}, D.~G., {et~al.} 2000, \aj, 120, 1579

\bibitem[{{Yuan} {et~al.}(2013){Yuan}, {Liu}, \& {Xiang}}]{Yuan2013}
{Yuan}, H.~B., {Liu}, X.~W., \& {Xiang}, M.~S. 2013, \mnras, 430, 2188

\bibitem[{Yuan {et~al.}(2014)Yuan, Liu, Xiang, Huo, Zhang, Huang, \&
  Zhang}]{Yuan2013b}
Yuan, H.-B., Liu, X.-W., Xiang, M.-S., Huo, Z.-Y., Zhang, H.-H., Huang, Y., \&
  Zhang, H.-W. 2014, in Feltzing S., Zhao G., Walton N., Whitelock P., eds,
  Proc. IAU Symp. 298, Setting the scene for Gaia and LAMOST, Cambridge
  University Press, pp. 240-245, preprint (arXiv: 1306.5614)

\bibitem[{{Zhang} {et~al.}(2013){Zhang}, {Liu}, {Yuan}, {Zhao}, {Yao}, {Zhang},
  \& {Xiang}}]{Zhang2013}
{Zhang}, H.-H., {Liu}, X.-W., {Yuan}, H.-B., {Zhao}, H.-B., {Yao}, J.-S.,
  {Zhang}, H.-W., \& {Xiang}, M.-S. 2013, Research in Astronomy and
  Astrophysics, 13, 490

\bibitem[{{Zhang} {et~al.}(2014){Zhang}, {Liu}, {Yuan}, {Zhao}, {Yao}, {Zhang},
  \& {Huang}}]{Zhang2014}
{Zhang}, H.-H., {Liu}, X.-W., {Yuan}, H.-B., {Zhao}, H.-B., {Yao}, J.-S.,
  {Zhang}, H.-W.~{Xiang}, M.-S., \& {Huang}, Y. 2014, Research in Astronomy and
  Astrophysics, 14, 456

\end{thebibliography}

\end{document}